# Accurate Force Field Parameters and pH Resolved Surface Models for Hydroxyapatite to Understand Structure, Mechanics, Hydration, and Biological Interfaces

by

Tzu-Jen Lin[1] and Hendrik Heinz[1,2*]

Department of Polymer Engineering, University of Akron, Akron, OH 44325, USA

Department of Chemical and Biological Engineering, University of Colorado-Boulder, Boulder, CO 80309, USA

[*] Corresponding author: hendrik.heinz@colorado.edu

**Abstract**

Mineralization of bone and teeth involves interactions between biomolecules and hydroxyapatite. Associated complex interfaces and processes remain difficult to analyze at the 1 to 100 nm scale using current laboratory techniques, and prior models for atomistic simulations are limited in the representation of chemical bonding, surface chemistry, and interfacial interactions. This work introduces an accurate force field along with pH-resolved surface models for hydroxyapatite to represent chemical bonding, structural, surface, interfacial, and mechanical properties in quantitative agreement with experiment. The accuracy is orders of magnitude higher in comparison to earlier models to facilitate quantitative monitoring of inorganic-biological assembly. The force field is integrated into the CHARMM, AMBER, OPLS-AA, PCFF, and INTERFACE force fields to enable realistic simulations of apatite-biological systems of any composition and ionic strength. Specifically, the parameters reproduce lattice constants (<0.5% deviation), IR spectrum, cleavage energies, immersion energies in water (<5% deviation), and elastic constants (<10% deviation) of hydroxyapatite in comparison to experiment. Interactions between mineral, water, and organic compounds are represented by standard combination rules in the force field without additional adjustable parameters and shown to achieve quantitative accuracy. Surface models for common (001), (010), (020), (101) facets and nanocrystals are introduced as a function of pH on the basis of extensive experimental data. New insight into surface and immersion energies, the structure of aqueous interfaces, density profiles, and superficial dissolution is described. Mechanisms of specific binding of peptides, drugs, and mineralization can be analyzed and the force field is extensible to substituted and defective apatites as well as to other calcium phosphate phases.

# 1. Introduction

Apatites and related calcium minerals are of great importance for human health due to the presence in bone and teeth, as well as their role in maladies such as osteoporosis, caries and coronary calcification that affect the quality of life of billions of people.[1-24] Better understanding and engineering of the complex surface chemistry and assembly of apatite nanostructures embedded in biological environments from the atomic to microstructural scale could facilitate major discoveries. Current experimental techniques, however, have difficulties to allow visualization and thermodynamic analysis at the 1 to 100 nm scale and would benefit from realistic molecular simulations in comparison with laboratory tests and clinical studies.

Hydroxyapatite (HAP), $Ca_{10}(PO_4)_6(OH)_2$ is a major mineral phase in human bone and teeth embedded in a matrix of collagen and other biomolecules.[1, 2, 4, 7, 12, 25-33] Substitutions leading to alternative composition are common such as partial replacement of calcium ions by magnesium ions, hydroxide ions by fluoride ions, or phosphate groups by carbonate groups. Biomineralization usually involves biopolymers and small molecules which control of the spatial orientation and pattern of crystal growth in a cell, the composition and type of mineral phases, as well as the development of facets, shape, and crystallite size.[1, 2, 25] Specific interactions between apatite surfaces and proteins are known to involve electrostatic interactions,[34-38] hydrogen bonds, hydrophobic interactions, and conformation effects.[36, 38]

Knowledge of such interactions can provide understanding of the affinity of biomolecules to HAP surfaces. The elucidation of the relative importance of interactions and mineralization mechanisms has yet remained an enigma since quantitative data from experimental measurements remain limited, adequate simulation tools were not available, and the phase space of possible environments and structures has virtually no limits even at the nanometer scale. From a chemical perspective, the fundamental mechanisms of recognition are expected to be comparable to those for biomolecule adsorption on silica, cement minerals, and other oxidic surfaces with similar density of ionic groups per unit area.[39-43] For example, ion pairing, hydrogen bonds, hydrophobic depletion interactions, as well as conformation effects have been described in detail. Specific recognition and crystallization properties of apatites are likely associated with the distinctive surface structure and surface chemistry as a function of crystallographic facets, facet size,[44] defects, as well as changes in pH and other solution conditions.[39]

Details of the structure and composition of hydroxyapatite surfaces have been obtained by solubility measurements, NMR spectroscopy, titration, water vapor adsorption, IR, and AFM studies that indicate notable differences to the structure of the bulk mineral.[45-61] At the same time, uncertainties related to the use of polycrystalline samples versus single crystals, well-defined even surfaces versus imperfect hydrated surfaces, the type of displayed facets, defects, and specific details on biomolecular adsorption also need to be acknowledged. For example, some reports indicate that peptides with a high proportion of glutamic and aspartic acids favor adsorption; yet other reports highlight the attraction of peptides with few or no acidic amino acid groups to apatite-based materials.[9, 21, 23, 36, 62-64] Studies using phage display found that the peptide sequence SVSVGMKPSPRP has a high affinity to HAP surfaces.[22, 65] By means of recent simulations it could be shown that the attraction of the amino acid motif SVSV to the hydroxyapatite prismatic plane is strongly pH dependent with significant adsorption at pH values found in bone (pH ~ 5) while a different set of ionic residues preferentially adsorbs at higher pH values (pH ~ 10), beginning to clarify seemingly contradictory experimental results.[39, 66]

Computer simulations hold promise to complement laboratory methods on an equal footing. The influence of specific crystallographic facets, nanocrystal size and shape, pH, temperature, and molecular concentration on biological recognition and assembly of nanostructures has been studied in atomic detail for similar systems.[39, 41, 42, 44, 67-72] Reliable force fields and surface models for apatites can eventually reveal contributions of different binding mechanisms of biomolecules to HAP surfaces and nanostructures, thereby advancing the understanding of crystal growth and dissolution. As an ultimate goal, molecular-level signatures of diseases such as osteoporosis, caries, and artery calcification can be better understood aided by computer simulations to support medical treatments and preventive care.

This work makes a first step towards such aims through the introduction of an accurate force field and a surface model database for hydroxyapatite. The proposed model takes into account the chemistry and stoichiometry of hydroxyapatite, includes the interpretation of all parameters, and reproduces bulk and surfaces properties quantitatively. Facet-specific as well as pH-specific interactions with water will be quantified, and follow-on contributions will explain selective interactions with peptides and drug molecules.[66] The hydroxyapatite model is integrated into the INTERFACE force field and easily extensible for alternate stoichiometry using parameters for other ions and compounds.[73]

The outline is as follows. In section 2, available simulation methods, limitations of prior models, and new capabilities are described. In sections 3 and 4, the force field parameters and surface models of hydroxyapatite are introduced including different facets and pH conditions. Evaluation of bulk properties, surface properties upon cleavage, as well as properties of aqueous hydroxyapatite interfaces follows in sections 5, 6, and 7, including a brief example of facet and pH-dependent peptide recognition. Conclusions are presented in section 8, and complete simulation protocols are provided in the Supporting Information. This work also covers a broad range of of experimental references to back up the development of models, as well as prior related computational references, which are discussed in detail in the respective sections.

## 2. Prior State-of-the-Art and New Capabilities

Simulation methods at length scales up to 1000 nm include ab-initio molecular dynamics, classical molecular dynamics (MD), and hybrid methods such as QM/MM.[73] Quantum mechanical calculations, including density functional theory (DFT), yield detailed information about the electronic structure at the scale of molecules and small molecular assemblies. Models and simulations in the condensed phase may include hundreds of atoms and picosecond time scales, such as several water molecules and short peptides in ab-initio molecular dynamics simulations on HAP surfaces in vacuum.[18, 74-79] The inclusion of explicit solvent, realistic ionic concentrations, pH, and sampling of the conformations of biopolymers on nanostructures with thousands to millions of atoms is, however, difficult. The strength of DFT and higher level electronic structure methods lies primarily in the analysis of local electronic structure, orbital energies, spatially oriented chemical bonding, and reactivity. DFT calculations can have significant uncertainties in computed surface tensions and binding energies in comparison to experimental measurements (up to 50%),[73, 80-82] and the large number of adjustable parameters in density functionals make it difficult to establish the relationship to simulation results.[83-86]

Classical all-atom molecular dynamics methods can use chemically realistic models up to length scales of 100 nanometers and simulation times up to microseconds. It is feasible to analyze differences in biomolecular binding as a function of crystallite size, facets, pH, concentration, temperature, and other solution conditions.[39, 41, 42, 44, 66-72, 87, 88] In particular, chemically consistent force fields for inorganic materials such as CHARMM-INTERFACE reproduce interfacial properties with only 0% to 10% uncertainty relative to experiment.[73] For example, molecular recognition mechanisms on silicates, aluminates, and metals have been identified using computation and experiment, including facet specific trend and correlations with nanocrystal shape and yield.[89, 90] Similarly good force field parameters and realistic atomistic models of hydroxyapatite have not been available and could equally predict interactions with biomolecules.[39, 66] The aim of this contribution is the description of suitable parameters and surface models in the context of available chemical knowledge, experimental reference data, and theory to provide new insight into surface, aqueous, and biological interfacial properties.

Some force fields for apatites have been proposed earlier. In general, chemical bonding and surface properties are only very approximately represented and surface models of realistic chemistry and pH have not been used at all. The INTERFACE protocol has shown that an accurate reflection of chemical bonding, dipolar interactions, and cohesion by means of atomic charges and Lennard-Jones parameters is critical for a chemical interpretation of force field parameters and improves the accuracy of predictions between one and two orders of magnitude. (1) A major drawback in prior apatite models has been the misrepresentation of ionic versus covalent contributions to bonding via atomic charges, then leading to arbitrary fitting of the remaining parameters to some properties and large unnecessary errors in others.[91] Examples are apatite force fields fitted to density and mechanical properties with errors of several 100% in surface and hydration energies that cannot be used to study inorganic-biological assembly.[92-97] (2) Similarly, Lennard-Jones parameters represent repulsive and attractive van der Waals interactions and the transcription from generic force fields such as UFF without interpretation, validation, and necessary modification results in unchecked energies and highly uncertain interfacial properties.[92-95, 98, 99] (3) As mentioned above, prior computational studies (except refs. [39, 66, 73]) have not regarded the pH dependence of the surface chemistry of hydroxyapatite (Figure 1).[92-100] These studies, including DFT and MD methods, assume freshly cleaved, non-hydrated hydroxyapatite surfaces at unrealistic pH conditions of approximately 15 that correspond to immediate cell death. We will show in this work that changes in pH profoundly transform the chemical nature of apatite surfaces and the interactions with water and biological molecules, similar to other pH responsive surfaces.[39] For example, prior studies on silica have demonstrated that even small changes in surface chemistry such as by two pH units alter the binding mechanism and reduce the sequence similarity of attracted peptides drastically, which is why inclusion of proper surface chemistry is essential.[39-41, 101, 102] (4) Earlier force field parameters for hydroxyapatite, for example, refs. [96, 100], also lack compatible with widely used force fields for biopolymers such as CHARMM,[103] AMBER,[104] GROMACS,[105] PCFF,[106] DREIDING,[107] and CVFF[108] due to a different energy function. Additional parameters are then necessary to describe the interaction between hydroxyapatite, solvents, and proteins,[109-111] which are not necessary with the proposed thermodynamically consistent hydroxyapatite force field here.

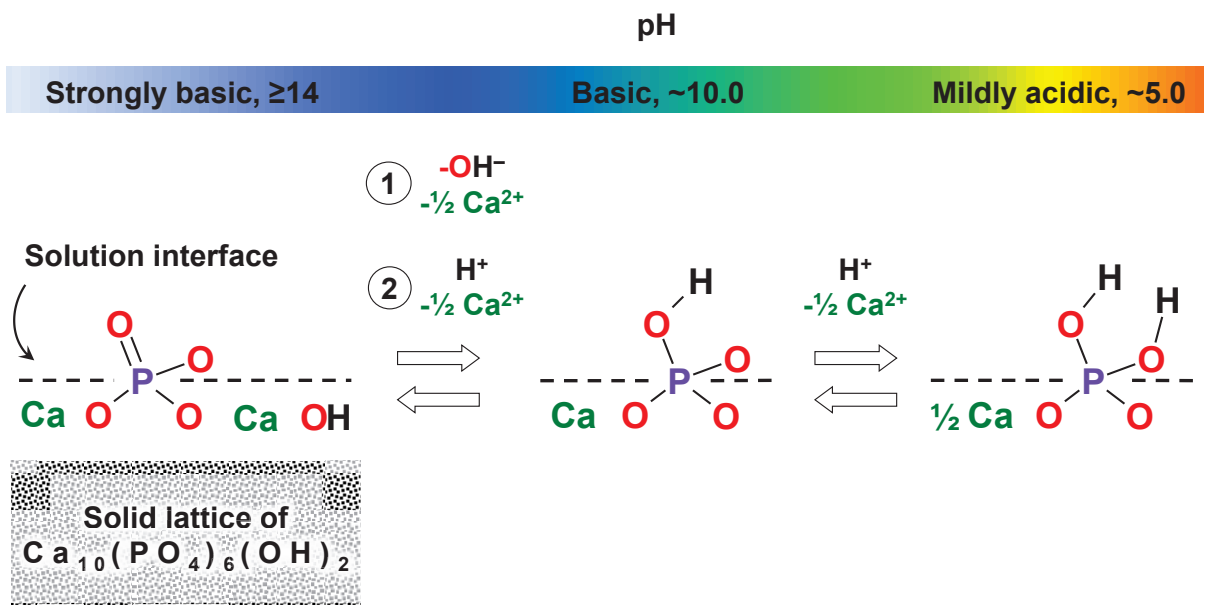

**Figure 1.** Schematic representation of HAP and the construction of surface models at different pH values. Interactions with water at pH values lower than 14 cause protonation and leaching of superficial hydroxide ions along with a stoichiometric amount of calcium ions. Increasing protonation of superficial phosphate ions leads to the formation of hydrogen phosphate and removal of a stoichiometric amount of calcium ions. At pH values lower than 10, further protonation of hydrogen phosphate to dihydrogen phosphate and removal of further calcium ions occurs. Surface models for a specific pH value can be customized by corresponding stoichiometric changes to pristine hydroxyapatite surfaces and associated redistribution of atomic charges keeping overall charge neutrality. Changes in crystal phase can also be implemented, for example, to octacalcium phosphate at low pH values.

The limitations are resolved by the present force field and surface models as follows. (1) Atomic charges represent chemical bonding in balance with available experimental data on electron deformation densities, similar compounds across the periodic table, chemical reactivity, theoretical foundations in the Extended Born model, and quantum chemistry.[73, 91] (2) The chosen Lennard-Jones parameters are accompanied by a physical explanation in the context of similar compounds and validated against densities, surface energies, and hydration energies. (3) The surface models take into account common (001), (010), (020), and (101) facets of hydroxyapatite nanocrystals (Figure 2) at protonation states across a range of pH values. (4) The parameters are integrated into the INTERFACE force field and compatible with the force fields CHARMM, AMBER, GROMACS, PCFF, CVFF, and OPLS-AA, including standard water models such as SPC and TIP3P. Additional parameters to simulate the inorganic-organic interfaces are not needed. (5) Validation includes the density, IR spectrum, mechanical properties, surface properties, and interfacial properties with water including the range of pH values in which apatite is stable (3 to 15). Computed properties from molecular dynamics simulations are in excellent agreement with experimental data, surpassing prior available models up to two orders of magnitude in accuracy.

The aim is quantitative analysis of an unlimited number of apatite surfaces and nanocrystals in combination with water, proteins, DNA, and organic molecules under realistic solution conditions. In this work, an analysis of facet-specific cleavage energies, hydration energies, and of the structure of the HAP-water interface as a function of pH is presented that reveals the impact of solution conditions on electric double layers and surface properties. The effect of crystallographic facets and pH on the selective binding of HAP-binding peptides (SVSVGGK) and drug molecules for osteoporosis will be described separately in detail (see preview in ref. [39, 66]).

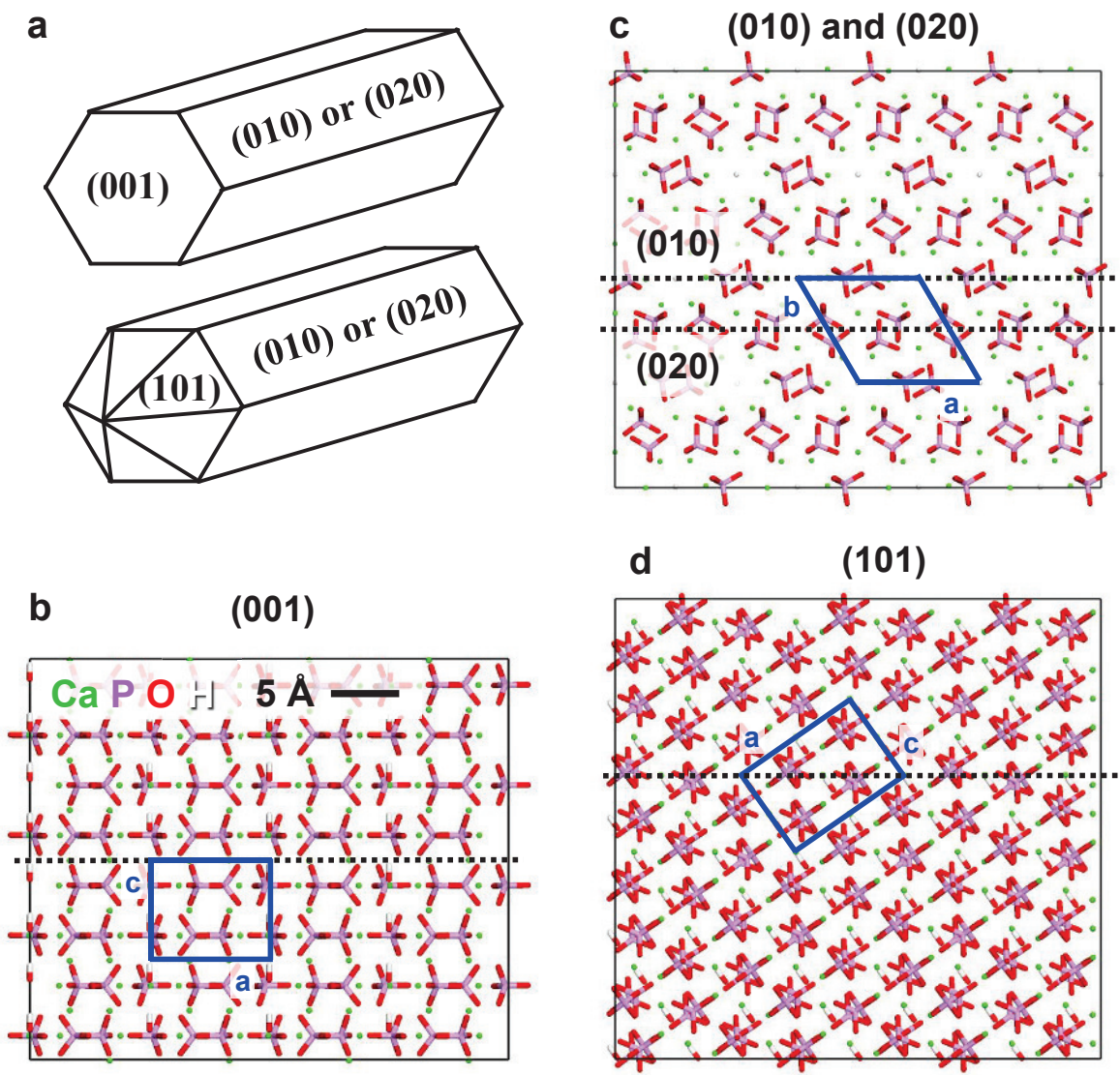

**Figure 2.** Nanocrystal shape and common cleavage planes of hydroxyapatite. (a) Schematic of the rod-like shape of HAP nanocrystals. (b) The basal plane (001). (c) The prismatic planes (010) and (020), respectively. (d) The (101) plane. Projections of the unit cell are shown in blue, and cleavage planes are shown as dashed lines in side view. Models of cleaved surfaces are prepared by an equal 50/50 distribution of ions on both sides of the cleavage plane followed by thermal relaxation and energy minimization.

## 3. Force Field Parameters and Surface Models

**3.1. Choice of Energy Expression.** The hydroxyapatite parameters extend the INTERFACE force field[73] following a series of developments for clay minerals, silica, cement minerals, aluminates, metals, and calcium sulfates.[42, 67, 71, 82, 112] The HAP parameters are compatible with the force fields PCFF, COMPASS, GROMACS, CVFF, DREIDING, CHARMM, AMBER, and OPLS-AA that cover solvents, organic molecules, and biopolymers. Broad compatibility supports the mission of the INTERFACE force field to unify materials oriented and biomolecular simulations in one single and accurate platform. The functional form of PCFF and COMPASS force fields consists of a quadratic potential for bond stretching and angle bending, a Coulomb potential, and a 9-6 Lennard-Jones potential using Waldman-Hagler combination rules between different atom types $i$ and $j$ (equation 1):

$$E_{total} = \sum_{bonds} K_{r,ij}(r_{ij} - r_{0,ij})^2 + \sum_{angles} K_{\theta,ijk}(\theta_{ijk} - \theta_{0,ijk})^2 + \sum_{ij\,nonbonded} \frac{q_i q_j}{4\pi\varepsilon_0 r_{ij}}$$
$$+ \sum_{ij\,nonbonded} \varepsilon_{ij}\left[ 2\left(\frac{\sigma_{ij}}{r_{ij}}\right)^9 - 3\left(\frac{\sigma_{ij}}{r_{ij}}\right)^6 \right] \quad (1)$$

The original energy expression comprises additional terms for higher order bond and angle stretching (cubic and quartic), torsion, out-of-plane, as well as cross-terms. However, such terms are not required for minerals and set to zero;[113] their presence for other compounds such as organic molecules and biopolymers does not affect the performance for apatite. Dihedral angles (torsions) can be found in O-P-O-H groups in monohydrogen or dihydrogen phosphate due to the terminal hydrogen atoms, however, these dihedral angles are reasonably described by non-bonded interactions and corresponding torsion terms are set to zero. The simplicity of the energy expression eases transferability of PCFF parameters to CVFF, GROMACS, CHARMM, DREIDING, AMBER, and OPLS-AA.

The force fields CVFF, CHARMM, GROMACS, DREIDING, AMBER, and OPLS-AA use a 12-6 Lennard-Jones potential instead of a 9-6 Lennard-Jones potential (equation 2):

$$E_{total} = \sum_{bonds} K_{r,ij}(r_{ij} - r_{0,ij})^2 + \sum_{angles} K_{\theta,ijk}(\theta_{ijk} - \theta_{0,ijk})^2 + \sum_{ij\, nonbonded} \frac{q_i q_j}{4\pi\varepsilon_0 r_{ij}}$$
$$+ \sum_{ij\, nonbonded} \varepsilon_{ij}\left[\left(\frac{\sigma_{ij}}{r_{ij}}\right)^{12} - 2\left(\frac{\sigma_{ij}}{r_{ij}}\right)^{6}\right] \quad (2)$$

The 12-6 force fields are very similar but still differ in combination rules for LJ parameters and in scaling factors for nonbond interactions between 1,4 bonded atoms. In CVFF, DREIDING, and OPLS-AA, geometric combination rules are used for $\sigma_{ij}$ and $\varepsilon_{ij}$; arithmetic combination rules are employed in CHARMM and AMBER for $\sigma_{ij}$ ($\varepsilon_{ij}$ is geometric in both). Scaling factors for nonbond interactions between 1,4 bonded atoms modify the influence of atomic charges and LJ parameters on bonded parameters (see ref. [71]). The scaling factors are 1.0 overall in CVFF, CHARMM, DREIDING, and typically also in GROMACS; 5/6 for Coulomb interactions and 0.5 for LJ interactions in AMBER, 0.5 for Coulomb interactions and 0.5 for LJ interactions in OPLS-AA. Nonbond interactions between 1,2 and 1,3 bonded atoms are excluded in all force fields (scaling factor of zero). The parameters for hydroxyapatite are the same for all force fields using a 12-6 Lennard-Jones potential regardless of these scaling conventions because 1,4 bonded atoms are only present in protonated form (O-P-O-H) and the impact is negligible.

**3.2. Rationale for Parameter Derivation.** The development of force field parameters follows the INTERFACE approach (Figure 3).[73] Key steps are (1) the definition of atom types, (2) the analysis of atomic charges, (3) the assignment of bonded terms and Lennard-Jones parameters, (4) successive validation and iteration of structural and energetic properties. The procedure has similarities to Rietveld refinement of X-ray data. Added benefits are (1) the representation of chemical bonding by atomic charges, (2) access to thermodynamic data and surface properties in the time domain, (3) applicability of the same parameters to an unlimited number of interfaces with inorganic, organic, and biomolecular compounds, as well as to chemically modified apatites.

Hydroxyapatite contains six distinct atom types, namely $Ca^{2+}$ ions, P, O, and H atoms in phosphate and hydrogenphosphate, as well as O and H atoms in hydroxide ions. The analysis of chemical bonding by atomic charges shows that only Ca is predominantly ionic, all other atoms are predominantly covalent. Therefore, bonded terms are included for all nearest neighbor atoms except Ca, and nonbonded terms are included between all pairs of atoms (under observation of scaling rules for 1,2, 1,3, and 1,4 bonded atoms).

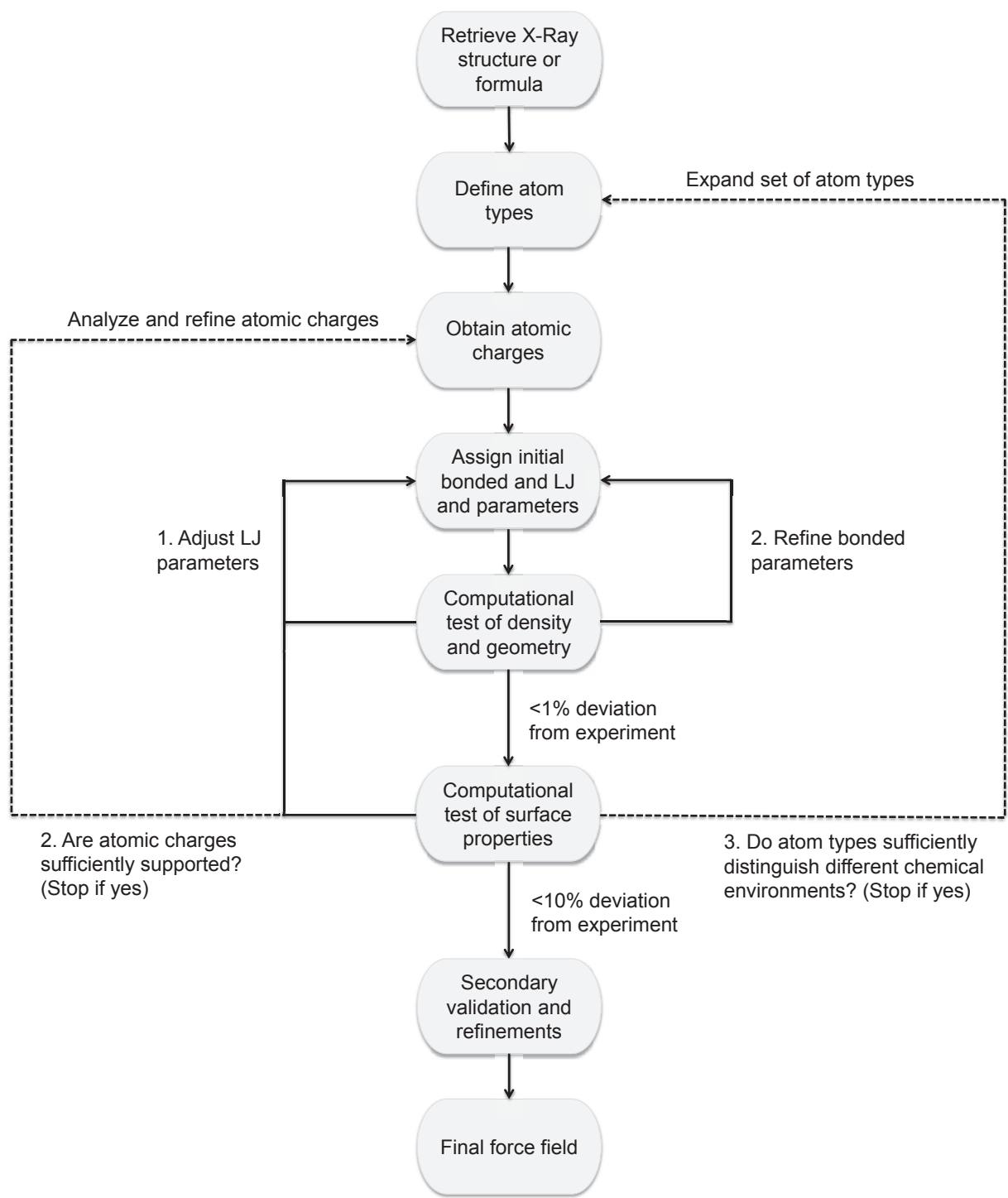

**Figure 3.** Flow chart for the parameter development for hydroxyapatite in the INTERFACE force field (from ref. [73]). The procedure is universal for compounds across the periodic table.

**3.3. Atomic Charges.** Atomic charges approximate the distribution of valence electrons (polarity of atoms in molecules).[91] They are used to quantify the balance of covalent bonding versus ionic bonding, thereby representing the nature of the chemical bond, and play a central role for reliable simulation outcomes. Already modest ionic character and atomic charges, for example, in silica and clay minerals, lead to dominance of Coulomb energy over bond, angle, and torsion energies.[40, 41, 67, 114] Atomic charges are also conceptually different from formal charges that help in bookkeeping of valence electrons and redox states. As explained earlier, formal charges are not a good instrument to explain the nature of chemical bonding and to represent atomic charges in force fields.[91] Formal charges only correlate with atomic charges for the elements themselves with zero charge, associated with covalent or metallic bonding, as well as for some alkali and earth alkali halides that exhibit full ionic bonding. Bonded terms are included in the force field whenever atomic charges for a given pair of neighbor atoms amount to less than half the formal charges, and when the interatomic distance is within the expected range for a predominantly covalent bond. Nonbonded terms are included among all atoms.

Atomic charges for classical simulations are specified in best convergence when multiple indicators are taken into consideration (Heinz charges).[91] These indicators include atomic charges according to electron deformation densities from X-ray data, dipole moments measured by spectroscopy, the extended Born model, chemical reactivity data, melting points, boiling points, solubility data, and any other readily available chemical and physical data related to polarity. Charges derived from *ab initio* studies are generally not suited to parameterize force fields due to dependence on method and large scatter.[91, 115] On the contrary, Heinz charges *intend* to measure chemical bonding and electronic structure in a simple manner, are accessible for any compounds across the periodic table, and achieve high accuracy when used in numerical computation. They are inherently suitable for the initial analysis of chemical reactivity and design of reactive force fields.

The atomic charge of calcium in hydroxyapatite was chosen as +1.5±0.1e consistent with the minerals tricalcium silicate,[42] tricalcium aluminate,[71] tobermorites, and gypsum that are chemically similar.[73] Si and S are the left and right neighbors of P in the periodic table, and cleavage energies as well as solid-water interfacial tensions of corresponding calcium sulfates and silicates were previously computed in agreement with experiment using a calcium charge of +1.5$e$. The coordination number of calcium ions, the type of ions present in these minerals, and melting points are also similar to hydroxyapatite.[61] The Ca charge +1.5e is notably below +2.0$e$ and clearly shows residual covalent character of bonding between calcium ions and coordinating oxygen atoms, which has been similarly reported for other calcium compounds according to deformation electron densities from X-ray data.[73, 91] Cases with confirmed formal charges for $Ca^{2+}$ (+2.0e) are solid $CaF_2$ related to the high electron affinity of fluorine as well as $Ca^{2+}$ ions in aqueous solution due to the distribution of negative charge across water molecules in the hydration shell.

The atomic charge of the phosphorus atom in HAP was derived as +1.0±0.1$e$ from the X-ray deformation density of aluminum phosphate.[116] This phosphorus charge in tetrahedral oxygen coordination is similar to that of silicon in tetrahedral oxygen coordination (+1.1±0.1e), supported by a slightly higher sum of atomization energy, first ionization potential, and partial second ionization potential (1.33 vs 1.24 MJ/mol at a charge of +1.0e).[91]

Hydroxide ions in hydroxyapatite, and analogously fluoride ions in fluoroapatite, are located in channels surrounded by calcium ions.[117-119] The net atomic charge of the hydroxide ion in the mineral was set as -0.9±0.1$e$, i.e., slightly below -1.0$e$, due to residual covalent bonding known in $Ca(OH)_2$.[42, 71, 91, 120] The atomic charges within the hydroxide ion are +0.2$e$ for the hydrogen atom and -1.1$e$ for the oxygen atom. As a further rationale, one may assume a water molecule (SPC, TIP3P) with atomic charges of -0.82$e$ and +0.41$e$ for O and H, followed by abstraction of a proton (here +0.9$e$ only) and redistribution of the excess negative charge of -0.49$e$ over O and H with a higher fraction on the more electronegative O.

To satisfy charge neutrality of hydroxyapatite, $Ca_{10}(PO_4)_6(OH)_2$, the atomic charge of the oxygen atom in the phosphate is then -0.8$e$. This result is consistent with atomic charges for oxygen atoms in clay minerals and other oxides in a range of -0.5$e$ to -0.9$e$, as well as anknown minimum of the the electron affinity of oxygen near -0.6$e$ to -0.7$e$ and.[73, 91] The negative charge of -0.8$e$ on the oxygen atoms in phosphate is at the higher end of this spectrum due to the overall negative charge of the non-protonated phosphate ions. Upon protonation to hydrogenphosphates at lower pH, the difference between incoming proton charge and leaving calcium charge distributes over the oxygen atoms and decreases the oxygen partial charge in a range from -0.8$e$ to -0.6$e$ (see section on surface models for details).

Table 1. Non-bonded parameters for bulk HAP and protonated surfaces containing monohydrogen phosphate as well as dihydrogen phosphate. The parameters are listed for PCFF, CFF, COMPASS (9-6 Lennard-Jones potential), GROMACS, AMBER, CHARMM, DREIDING, CVFF, and OPLS-AA (12-6 Lennard-Jones potential).[a]

| Atom type | Charge (e) | 9-6 LJ parameters (PCFF, CFF, COMPASS) | | 12-6 LJ parameters (CHARMM, CVFF, AMBER, GROMACS, DREIDING, OPLS-AA) | |
|---|---|---|---|---|---|
| | | $\sigma_0$ (Å) | $\varepsilon_0$ (kcal/mol) | $\sigma_0$ (Å) | $\varepsilon_0$ (kcal/mol) |
| **Bulk hydroxyapatite** | | | | | |
| Ca | 1.5 | 3.55 | 0.240 | 3.30 | 0.130 |
| P | 1.0 | 4.50 | 0.250 | 4.30 | 0.280 |
| O in $PO_4^{3-}$ | -0.8 | 3.50 | 0.055 | 3.40 | 0.070 |
| O in $OH^-$ | -1.1 | 3.80 | 0.080 | 3.70 | 0.080 |
| H in $OH^-$ | 0.2 | 1.098 | 0.013 | 0.0001 | 0 |
| **Hydrated and protonated surface** | | | | | |
| O in $(H\mathbf{O})_nPO_{4-n}^{(3-n)-}$ | -0.65 | same as O in $PO_4^{3-}$ | | | |
| O in $(HO)_nP\mathbf{O}_{4-n}^{(3-n)-}$ | -0.6 to -0.8[b] | same as O in $PO_4^{3-}$ | | | |
| H in $\mathbf{H}PO_4^{2-}$ /$\mathbf{H}_2PO_4^-$ | 0.4 | 1.098 | 0.013 | 0.0001 | 0 |

[a] The 12-6 Lennard-Jones parameters are identical for AMBER and CHARMM due to the absence of 1,4 bonded atoms and identical combination rules for $\sigma_0$ (arithmetic) and $\varepsilon_0$ (geometric). CVFF and OPLS-AA use geometric combinations of $\sigma_0$ that lead to practically identical results

without adjustments. [b] The atomic charge on the oxygen atoms in protonated phosphates depends on the degree of protonation and on the (h k l) facet due to different amount of calcium hydroxide leakage. See discussion of surface models in section 4.

**3.4. Bonded Parameters.** Bonded parameters of HAP include equilibrium bond lengths $r_{0,ij}$, angles $\theta_{0,ijk}$, harmonic bond stretching coefficients $K_{r,ij}$, and harmonic angle bending coefficients $K_{\theta,ijk}$ (equations (1) and (2)). The parameters were derived from experimental X-ray data and vibration spectra (Table 2). Equilibrium bond lengths of P-O bonds in phosphate and O-H bonds in hydroxide ions are 1.53 Å and 0.955 Å according to X-ray diffraction, respectively.[117-119] The parameters $r_{0,ij}$ were chosen 1.57 Å for P-O and 0.94 Å for O-H to reproduce experimental values with <1% deviation in NPT molecular dynamics simulation at 298 K. The small differences between the parameters $r_{0,ij}$ in the force field and the true equilibrium bond lengths in experiment are caused by the superposition of strong electrostatic interactions in hydroxyapatite that contract the P-O bond length and elongate the O-H bond length. The equilibrium bond length $r_{0,ij}$ of O-H groups in protonated phosphate was also chosen as 0.94 Å (Table 2).

The equilibrium angle $\theta_{0,ijk}$ of O-P-O bending was chosen as 109.47° consistent with XRD studies and sp$^3$ hybridization (Table 2).[117-119] The angle for the H-O-P bonds in protonated phosphate was assigned as 115° in analogy to parameters for clay minerals.[112] Average computed bond angles deviate less than 1% from XRD data in NPT molecular dynamics simulation under standard conditions.

The harmonic force constants for bond stretching $K_{r,ij}$ and angle bending $K_{\theta,ijk}$ were tuned to reproduce the experimental IR spectrum and agree within ±50 cm$^{-1}$ (Figure 4). Three major IR bands of hydroxyapatite can be observed, including O-H stretching at 3600 cm$^{-1}$, P-O stretching at 1100 cm$^{-1}$, and O-P-O bending at 600 cm$^{-1}$.[121-124] The IR band at 3600 cm$^{-1}$ originates from the bond stretching coefficient of O-H bonds in hydroxide ions and in hydrogenphosphate ($K_{r,ij}$ = 500 kcal/(mol · Å$^2$)). The P-O stretching band at 1100 cm$^{-1}$ is represented by the P-O stretching constant $K_{r,ij}$ = 430 kcal/(mol · Å$^2$). The P-O band is similar to Si-O and Al-O stretching bands in silicates and clay minerals at 1050 cm$^{-1}$, supported by the similar mass of Al, Si, and P atoms. Therefore, also the value of $K_{r,ij}$ resembles that of (Si, Al)-O stretching in layered silicates as described by Heinz et al.[112] Further analogies were found for the angle bending coefficients $K_{\theta,ijk}$ of tetrahedral O-P-O and O-Si-O bond angles. Isolated phosphate groups in hydroxyapatite are described with $K_{\theta,ijk}$ = 125 kcal/(mol · rad$^2$) and reproduce the experimentally observed IR band at 600 cm$^{-1}$ (Figure 4). The bending constant maintains the tetrahedral geometry of phosphate ions and was equally applied to protonated phosphate ions on the hydroxyapatite surface. In comparison, $K_{\theta,ijk}$ for O-(Al, Si)-O angles in clay minerals can reach up to 170 kcal/(mol · rad$^2$), which occurs due to the presence of an extended framework of covalent bonds that increases the reduced mass $\mu$ of vibrational subunits,[42, 67, 112] resulting in higher bending coefficients $k$ to yield the same vibration frequency ($\Box^2 = \Box/k$).

The H-O-P-O bond sequence also contains a dihedral angle. The dihedral potential was set to zero since the superimposed LJ potential describes dihedral preferences of hydrogen atoms in good approximation, and the absence of a torsion potential increases the compatibility among force fields with different functional forms (PCFF, CHARMM, DREIDING, GROMACS, AMBER, CVFF, OPLS-AA).

**Table 2.** Bonded parameters for HAP and protonated phosphates. The bonded parameters are identical for all energy expressions (equations 1 and 2).

| Bond type | $r_{0,ij}$ (Å) | $K_{r,ij}$ [kcal/(mol · Å$^2$)] |
|---|---|---|
| P–O bonds in $PO_4^{3-}$, $HPO_4^{2-}$, $H_2PO_4^-$ | 1.570 | 430 |
| O–H bonds in $HPO_4^{2-}$, $H_2PO_4^-$ and $OH^-$ | 0.940 | 500 |
| Angle type | $\theta_{0,ijk}$ (°) | $K_{\theta,ijk}$ [kcal/(mol · rad$^2$)] |
| O–P–O in $PO_4^{3-}$, $HPO_4^{2-}$, $H_2PO_4^-$ | 109.47 | 125 |
| P–O–H in $HPO_4^{2-}$, $H_2PO_4^-$ | 115.0 | 50 |
| Dihedral type | $\varphi_{0,ijkl}$ (°) | $K_{\varphi,ijkl}$ (kcal/mol) |
| H-O-P-O in $HPO_4^{2-}$, $H_2PO_4^-$ | 0 | 0 |

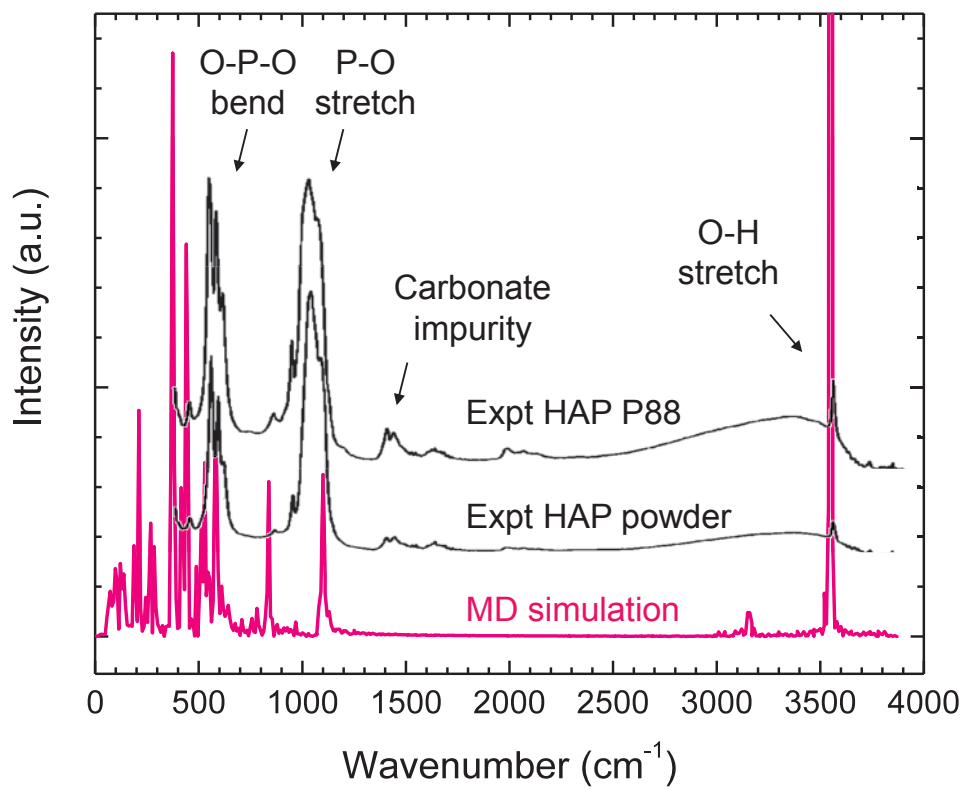

**Figure 4.** IR spectrum of hydroxyapatite in experiment and in molecular dynamics simulation (MD). Experimental FTIR spectra were recorded for HAP P88 (Plasma Biotal) and HAP powder (Merck), see ref. [123]. The computed spectrum equals a superposition of IR and Raman signals.

**3.5. Lennard-Jones Parameters.** Lennard-Jones parameters are essential to reproduce the density and surface properties.[67, 73, 112] The nonbond diameter $\sigma_0$ represents the van-der-Waals size of each atom type and is approximately known for all elements and ions across the periodic table.[61, 125] The actual values in the force field commonly fall within a ±10% range of the tabulated values depending on the chemical environment; larger values often serve to compensate attractive electrostatic interactions in significantly ionic systems. The primary function of $\sigma_0$ is the reproduction of the experimentally observed density. The parameter $\varepsilon_0$ correlates in first order approximation with the polarizability of each atom type. In highly charged systems, significantly increased values $\varepsilon_0$ also help counteract strong electrostatic attraction. High values of $\varepsilon_0$ then increase the slope of the repulsive part of the LJ potential, similar to a small increase in $\sigma_0$. In addition, the pairwise summation of LJ interactions introduces a dependence of $\varepsilon_0$ on the number of neighbor atoms per unit volume in otherwise comparable chemical environments. For example, $\varepsilon_0$ of oxygen atoms in a dense covalent oxide framework with a high number of nonbonded neighbors is smaller than $\varepsilon_0$ of oxygen in a molecular liquid such as water with comparable bond types and a lower number of nonbonded neighbors per unit volume. The validation of $\varepsilon_0$ ultimately requires the reproduction of a cleavage energy, hydration energy, or other surface property at 298 K, which typically leads to reliable predictions of several associated properties.[42, 73, 112] Such validation is detrimental as the classical Hamiltonian is then also calibrated for energies in addition to the calibration for structure.

Interpretations of non-bonded energies $\varepsilon_0$ were also previously discussed, for example, by Halgren,[126] Miller and Savchick,[127] Nagle,[128] and Rappe et al.,[129] although without reference to inorganic solids and without a discussion of the average number of nonbonded neighbors. The interpretation of $\varepsilon_0$ as a polarizability[126] is helpful to understand trends in the periodic table and assign initial values when the degree of ionic bonding is comparatively small. Accordingly, $\varepsilon_0$ increases for elements from top to bottom in each column of the periodic table, and from left to right in each row of the periodic table, as well as from positive charge towards negative charge on atoms in polar compounds. Exceptions are elemental metals that exhibit very high $\varepsilon_0$ due to extraordinary cohesion and lack of covalent bonds.[82] The final assignment of $\varepsilon_0$ requires validation of a cleavage energy, surface free energy, or hydration energy and is of equal importance as the assignment of atomic charges.

LJ parameters for the atom types in hydroxyapatite, as also the other parameters, are similar to those in silicates and gypsum as a consequence of similar chemical bonding and coordination environments (Table 2).[42, 67, 73] The van der Waals diameter $\sigma_0$ of calcium is 3.55 Å in the 9-6 LJ potential and 3.30 Å in the 12-6 potential, slightly larger than for a fully ionized $Ca^{2+}$ ion due to some covalent character in the minerals.[61, 125] $\sigma_0$ is thus larger than for a $Ca^{2+}$ ion in the CHARMM force field (2.734 Å) that reproduces the solvation energy in water but cannot procreate the hydroxyapatite crystal structure. Phosphorus is represented by a $\sigma_0$ value of 4.50 Å in the 9-6 LJ potential and 4.30 Å in the 12-6 LJ potential, similar to silicon in silicates. The value $\sigma_0$ is comparatively large due to the position of P in the 3rd row of the periodic table and the requirement of increased repulsion as a result of the significant positive charge. $\sigma_0$ values of oxygen atoms in phosphate are 3.50 Å and 3.40 Å in the 9-6 and 12-6 LJ potential, respectively, similar to oxygen in silicates. $\sigma_0$ values of oxygen atoms in hydroxide are somewhat larger at 3.80 Å and 3.70 Å, respectively, due to a higher negative charge and required repulsion (Table 1).

The non-bonded well depth $\varepsilon_0$ of calcium was set to 0.24 kcal/mol in PCFF and equivalently to 0.13 kcal/mol in CHARMM (Table 1). This comparatively large non-bonded well depth, especially in the 9-6 Lennard-Jones potential, strengthens the repulsive part of the potential to counterbalance strong Coulomb attraction. The dominance of Coulomb interactions would otherwise supersede the small expectation value of $\varepsilon_0$ for positively charged ions of low polarizability. The value $\varepsilon_0$ for phosphorus is 0.25 and 0.28 kcal/mol, respectively. Halgren's rule suggests $\varepsilon_0$ values in a range from 0.084 to 0.2846 kcal/mol for elements in the 3$^{rd}$ row of the periodic table; values at the high end help counterbalance ionic attraction.[126] The typical range of non-bonded well-depths $\varepsilon_0$ for oxygen as part of the second row the periodic table is 0.0218 to 0.084 kcal/mol. Values were chosen accordingly as 0.07 kcal/mol for oxygen atoms in phosphates, and a somewhat higher value $\varepsilon_0$ of 0.08 kcal/mol was assigned to hydroxide related to the higher charge. The values of $\varepsilon_0$ are generally higher than in extended covalent frameworks.[42, 67, 112]

The choice of $\varepsilon_0$ also depends on values of $\sigma_0$ and is not highly sensitive as long as the accurate reproduction of cell parameters and relative proportions to the remaining nonbond parameters are observed. For the transcription of parameters from the 9-6 LJ potential to the 12-6 LJ potential, van der Waals diameters $\sigma_0$ were reduced by approximately 5%. The well depth $\varepsilon_0$ in the 12-6 form tends to be larger than in the 9-6 form in case it represents dispersive interactions and compensates the stronger repulsion at shorter distance in the 12-6 Lennard-Jones potential.[112] However, the well depth $\varepsilon_0$ in the 12-6 form can be smaller in case it represents repulsion (e.g., for $Ca^{2+}$); which is then owed to the more repulsive nature of the 12-6 LJ potential. The final set of parameters was obtained upon thorough validation of density, cell geometry, as well as cleavage energy in over 500 individual tests.

## 4. Surface Models of Hydroxyapatite as a Function of pH

A critical aspect for modeling of apatite surfaces is the strong pH dependence of the surface chemistry. Realistic models make the difference between quantitative predictions and no predictions at all. A number of independent experimental studies explain changes in surface composition upon cleavage and hydration and are a great starting point to inform appropriate models.[45-61] The surface chemistry is strongly related to the distinct pK values of phosphoric acid of 2.15 ($pK_1$), 7.20 ($pK_2$), 12.32 ($pK_3$), as well as the pK value of water of 15.7 ($pK_W$) at 298 K.[61] The implementation in apatite models is an indispensable requirement for meaningful simulations of aqueous and biomolecular apatite interfaces.[39] Details of cleavage, hydration, and protonation of the surfaces and the representation in atomistic models are described in the following sections.

**4.1. Cleavage.** Cohesion in minerals is generally a result of covalent and non-covalent interatomic interactions. Covalent bonds tend to be stronger than nonbonded interactions such as ionic interactions, hydrogen bonds, dipolar, and van der Waals interactions. The number of possible cleavage planes is in principle unlimited while cleavage preferably occurs in directions with weaker nonbonded interactions (Figure 2). The most likely cleavage planes can often be identified by the analysis of distinct "layers" of cations and anions in the unit cell in a given direction[42, 130] and typically align with the direction of the cell vectors or simple combinations thereof. The crystal structure of hydroxyapatite involves alternating layers of positive and negative charges that are overall neutral within a repeat unit. Several cleavage planes of HAP can be considered as Type III surfaces according to a classification by Tasker,[131] i.e., charged and with a dipole moment in the repeat unit perpendicular to the surfaces. Surface models were created by equally dividing charged ions between the two newly created surfaces upon cleavage, followed by relaxation in molecular dynamics simulation to minimize local electric fields (Figures 2 and 5).[42, 71, 130, 132] Surface reconstruction leads to two low energy surfaces that are both charge-neutral.

Equilibrium cleavage planes were specifically created for the common (001), (010), (020), and (101) surfaces of hydroxyapatite (Figures 2 and 5). Upon cleavage of a (001) plane, a layer of calcium ions is equally divided between the two emerging (001) surfaces. The new positions of the superficial calcium ions were obtained by short MD simulations at high temperature.[42, 132] Cleavage of a (010) plane involved three kinds of ions: calcium, hydroxide, and phosphate ions, which are equally distributed among the two new surfaces. Cleavage leading to a (020) surface involved the equal distribution of phosphate ions only, and the (101) surface was obtained without the necessity to redistribute ions (Figure 2). Surface models of lowest energy were obtained through minimization of local electric fields and thermal annealing of the topmost atomic layers as previously described.[42, 130, 132] Rapid cleavage can also cause uneven distributions of ions in experiment, which leads to strong, distance-dependent attraction over a long range. ).[42, 130, 133] Models for non-equilibrium surfaces can be similarly derived but are not in focus here.

**4.2. Hydration and Influence of pH.** The unmodified cleaved surfaces of hydroxyapatite are hygroscopic and immediately attract water (or carbon dioxide) in air. The hydroxide ions then react to form a superficial hydrated layer of calcium hydroxide:

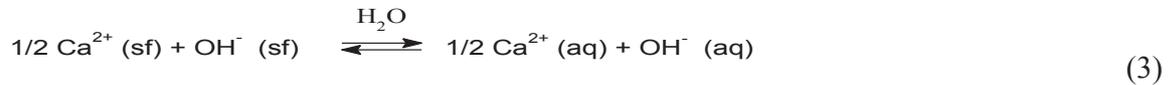

$$1/2\ Ca^{2+}\ (sf) + OH^-\ (sf) \underset{}{\overset{H_2O}{\rightleftharpoons}} 1/2\ Ca^{2+}\ (aq) + OH^-\ (aq) \quad (3)$$

$Ca^{2+}$ (sf) and $OH^-$ (sf) hereby denotes the ions at the top atomic layer of the surface and $Ca^{2+}$ (aq) and $OH^-$ (aq) the corresponding hydrated species formed in contact with water. The hydrated surfaces are only stable in solution at pH values above ~14 as the $pK_a$ value of water is 15.7.[61] Lower pH values cause neutralization of the superficial hydroxide ions and leaching of the hydroxyapatite surface. This process is driven by acids ($H^+A^-$) that are present in buffered solutions, including physiological conditions:

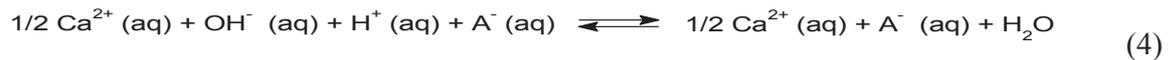

$$1/2\ Ca^{2+}\ (aq) + OH^-\ (aq) + H^+\ (aq) + A^-\ (aq) \rightleftharpoons 1/2\ Ca^{2+}\ (aq) + A^-\ (aq) + H_2O \quad (4)$$

Since pH values above pH 14 are practically uncommon in living organisms, the apatite surface is hardly ever present in chemically pure form. Hydration and protonation reactions initiate leaching of calcium hydroxide in the form of other calcium salts. The concentration of OH⁻ ions at the immediate surface of hydroxyapatite is very high (on the order of 1 mol/l), and due to the ion product of water $K_W = [H^+][OH^-] = 10^{-14} M^2$ any buffer system with a pH value less than 14 neutralizes the surface hydroxide (Figure 1).

The next process at pH values below 13 is the protonation of superficial phosphate ions to monohydrogenphosphate ions ($HPO_4^{2-}$):

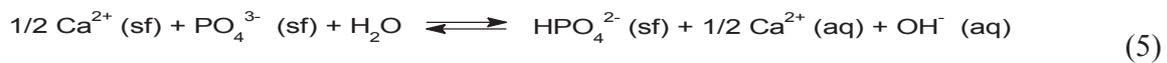

(5)

This process is related to the pK₃ value of 12.3 of phosphoric acid and causes further leaching of calcium hydroxide in subsurface layers. At a pH value of ~12.0, the surface layer consists approximately of a 50/50 mixture of phosphate and hydrogenphosphate ions. The net reactions and changes in surface chemistry at high pH are summarized in Figure 1.

A maximum amount of monohydrogenphosphate ions on the top surface layer near 100% is then expected at pH ~10.0, close to the average of pK₂ and pK₃ of 9.75. A further decrease in pH value leads to the protonation of monohydrogenphosphate to dihydrogenphosphate (H₂PO₄⁻):

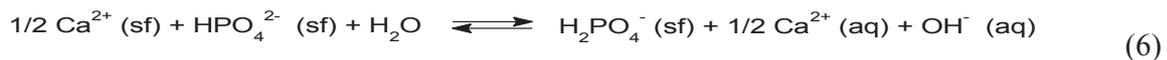

(6)

The associated pK₂ value of H₃PO₄ is 7.2 and leads to a 50/50 equilibrium mixture of hydrogenphosphate and dihydrogenphosphate on the surface at pH~7.5. This observation is consistent with the observed ratio of 31:69 at pH~7.0 by IR spectroscopy in experiment.[60] Finally, full coverage of the top surface layer with H₂PO₄⁻ ions is expected at a pH value of ~5.0 near the average of pK₁ and pK₂ of 4.7.

Below pH=5, surface termination may involve dihydrogen phosphate as well as phosphoric acid. Decreasing pH values also cause deeper penetration of water and protonation of phosphate ions up to many atomic layers beneath the surface atomic layer so that the chemical composition differs from "hydroxyapatite". The dissolution of hydroxide ions can lead to the precipitation of layered octacalcium phosphate; this process may begin already below pH 8.[58, 59] The ionization state of the surfaces also depends somewhat on the total ionic strength in solution and on the type of present ions.[40, 46, 134-136] In atomistic models, a distinction with a resolution of about 0.5 pH units appears to be sufficient for most purposes, especially in consideration of previous studies that assumed unrealistic pH values of 14 to examine processes under physiological conditions. [92-100]

**4.3. Implementation in Models.** Building models of apatite cleavage planes and nanocrystals for the different protonation stages involves cleavage, adjustments in the stoichiometry, and in atomic charges (Table 3). Stoichiometric adjustments are applied to the ions in the surface layer and in the desired number of subsurface layers; here only the surface layer was considered. The procedure is as follows: (1) implement changes in stoichiometry on the respective (h k l) cleavage plane according to the desired degree of hydration and protonation (Figure 1), (2) adjust atomic charges according to the formation of new bonds and departure of ions from the surface (e.g. adjustments from P-O bonds to P-O-H bonds when hydrogen atoms from water bind to phosphate ions and calcium and hydroxide ions are removed), (3) arrange cations and anions on the surface such that local electric fields and the overall energy are minimal using molecular dynamics. Thereby, the overall charge neutrality of the surfaces must be maintained[91] and associated charge increments/decrements must reasonably describe changes in chemical bonding and in the true electron density distribution.[67, 71]

In this manner, surface models and atomic charges can be determined and the process is designed to be simple, effective, and easy to customize. Upon changes in pH, only the charges of the oxygen atoms in phosphate ions require major modification while all other bonded and nonbonded force field parameters can be kept the same (Table 3). The protocol can be applied to prepare models of any (h k l) apatite surface and apatite nanocrystals. Customization for different degrees of water penetration and substitution of cations and anions, such as $Mg^{2+}$, $F^-$, $Cl^-$, $HCO_3^-$, and $CO_3^{2-}$ is also straightforward. The correlation between the proposed models and experimental observations is described in the following section.

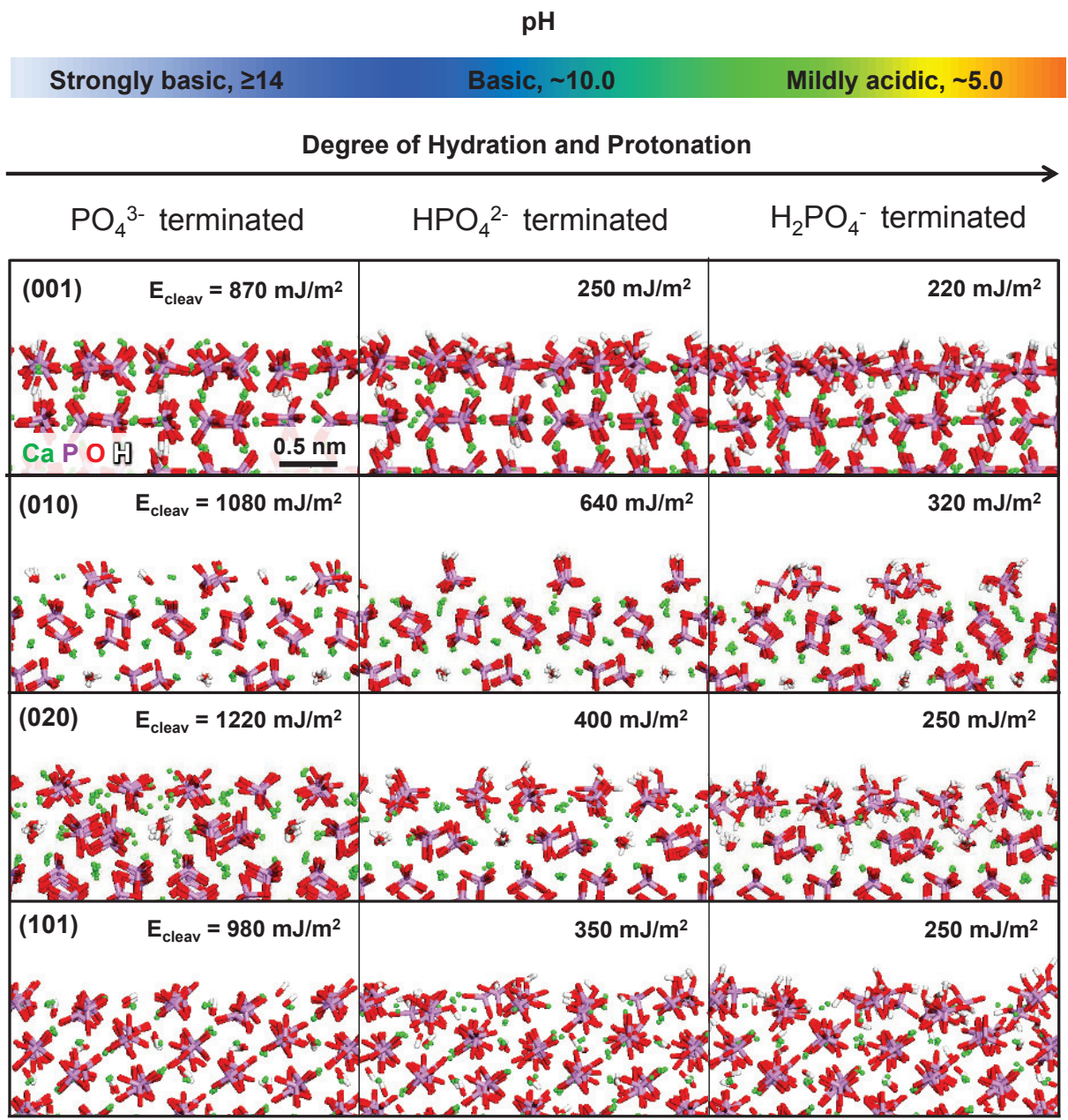

**Figure 5.** Side view of equilibrium model surfaces of hydroxyapatite and computed cleavage energies in vacuum at different pH values. (001), (010), (020), and (101) surfaces are shown. The left column shows the surfaces before hydration (pH > 14), the middle column shows the surfaces with monohydrogen phosphate termination (pH ~ 10), and the right column visualizes the surfaces with dihydrogen phosphate termination (pH ~ 5). For all facets, increasing surface hydration and protonation increase surface reconstruction, surface disorder, and decrease the cleavage energy. This trend correlates with decreased charge density per unit area as the pH value decreases from 14 to 5. The cleavage energies also show notable facet-specific detail. Protonation was limited to the topmost molecular layer in the models.

**Table 3.** Details of hydroxyapatite surface models. Atomic charges and stoichiometric changes are indicated as a function of protonation state and facet for a given surface area, assuming hydration and protonation in the topmost molecular layer.

|  | Facet | | | |
|---|---|---|---|---|
| Property | (001) | (010) | (020) | (101) |
| surface area (Å²) | 32.6214×37.6680 | 37.6678×34.3750 | 37.6678×34.3750 | 34.9787×32.6214 |
| **pH >14, PO₄³⁻ terminated, cleavage as-is** | | | | |
| atomic charges on phosphate ions[a] | 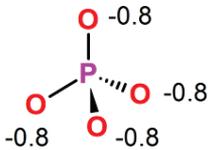 | 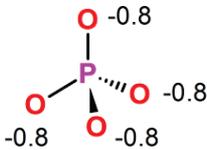 | 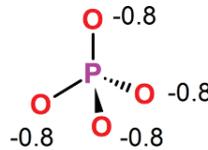 | 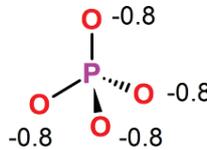 |
| **pH ~10, HPO₄²⁻ terminated** | | | | |
| # of hydroxide ions removed[b] | 16 | 20 | 0 | 12 |
| # of phosphate ions protonated[b] | 48 | 20 | 40 | 36 |
| # of calcium ions removed[b] | 32 | 20 | 20 | 24 |
| atomic charges on surface phosphate[a] | 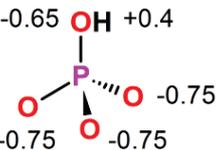 | 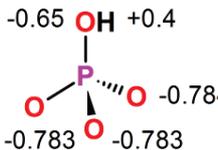 | 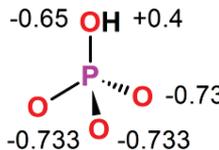 | 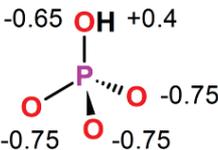 |
| **pH ~5, H₂PO₄⁻ terminated** | | | | |
| # of hydroxide ions removed[b] | 16 | 20 | 0 | 12 |
| # of phosphate ions protonated[b] | 48 | 20 | 40 | 36 |
| # of calcium ions removed[b] | 56 | 30 | 40 | 42 |

| atomic charges on surface phosphate[a] | 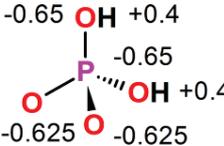 | 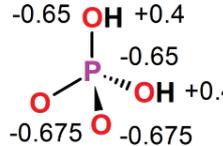 | 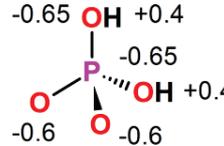 | 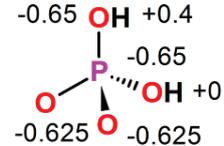 |
|---|---|---|---|---|

[a] The charge on the phosphorus atom was +1.0 $e$ for all pH values.

[b] At the topmost surface layer.

**4.4. Comparison with Experimental Data.** The proposed models are supported by laboratory measurements of acid-base equilibria, solubility products, adsorption isotherms, AFM, SEM, TEM, IR and NMR spectroscopy, XPS, and Auger electron spectroscopy.[45-61] The surface reactions described in equations (3) to (6) closely follow the protonation equilibria of water and phosphoric acid and are quantitatively in agreement with measurements. Measurements by IR spectroscopy have shown that the apatite surface consists of $HPO_4^{2-}$ and $H_2PO_4^-$ ions in a ratio of 31:69 at pH = 7 in water,[60] and are in agreement with an expected 50:50 ratio at pH = 7.20 that equals the p$K_2$ value for $H_2PO_4^-$ dissociation in water. Previous studies also show that the degree of ionization on crystal surfaces tends to be slightly lower in comparison to the same process in solution due to dense packing and increased ion-ion repulsion per unit area on the crystal surface.[67] Therefore, a 50:50 equilibrium between $HPO_4^{2-}$ and $H_2PO_4^-$ ions on the apatite surface may be expected at pH ~7.5 rather than at pH = p$K_2$ = 7.2.

Solubility measurements and NMR data have shown that the Ca:P ratio of 1.67 in native apatite ($Ca_{10}(PO_4)_6(OH)_2$) decreases on the surfaces to about 1.5 as a consequence of the altered stoichiometry upon hydration (Figure 1).[46, 49, 53] At lower pH, the Ca:P ratio even further decreases to below 1.0 consistent with the presence of $HPO_4^{2-}$ and $H_2PO_4^-$ ions ($CaHPO_4$ and $Ca(H_2PO_4)_2$, respectively).[58, 59] The exact Ca:P balance is thus dependent on crystal facet and pH, and represented in the proposed models (Table 3).

The depth of the hydrated and protonated layer was estimated to be about 1 nm in experiment[49] and could extend up to several nanometers at lower pH.[58, 59] The proposed surface models currently assume changes only in the first layer of phosphate ions for simplicity (~0.5 nm). The models represent most surface properties well and can be extended to any number of modified subsurface layers as needed (Table 3 and Figure 5). Further comparisons between experimental measurements and the proposed models are described in section S1 of the Supporting Information.

## 5. Validation of Structural, Vibrational, and Mechanical Properties

**5.1. Structural Properties.** Computed bond lengths, angles, and nonbonded interatomic distances in NPT simulations of bulk hydroxyapatite agree quantitatively with XRD measurements at room temperature.[118] Average bond lengths of P-O and O-H bonds are $1.54 \pm 0.01$ Å and $0.95 \pm 0.01$ Å, respectively, and O-P-O bond angles are $109.5° \pm 0.2°$. The deviation from experiment is less than 1%. Nonbonded interatomic distances between calcium ions and oxygen atoms in phosphate and hydroxide ions are between 2.4 and 2.6 Å in the simulation and depart less than 2% from X-ray data. Finally, computed lattice parameters in NPT simulation agree with measurements on average within ±0.5% (Table 5). Deviations slightly over 1% occur for the cell parameter $c$ using the 12-6 Lennard-Jones potential, related to stronger repulsive interactions. (Further customization of parameters to reduce the deviation appear not necessary.) The excellent reproduction of bonded and non-bonded interatomic distances, angles, and lattice constants is a testimony to chemically meaningful atomic charges, bonded parameters, and van-der Waals (LJ) parameters.

**Table 5.** Lattice parameters of hydroxyapatite in experiment and in NPT molecular dynamics simulation with the INTERFACE-PCFF and INTERFACE-CHARMM force field at 298.15 K and 101.3 kPa.

| Method | cell dim. | $a$ (Å) | $b$ (Å) | $c$ (Å) | $\alpha$ (º) | $\beta$ (º) | $\gamma$ (º) | Density (g/cm$^3$) |
|---|---|---|---|---|---|---|---|---|
| Experiment[a] | 2 × 2 × 1 | 18.83 | 18.83 | 13.75 | 90 | 90 | 120 | 3.160 |
| Force Field 9-6 | 2 × 2 × 1 | 18.82 | 18.81 | 13.75 | 90.00 | 90.00 | 120.02 | 3.167 |
| Force Field 12-6 | 2 × 2 × 1 | 18.78 | 18.77 | 13.95 | 90.00 | 90.00 | 119.99 | 3.135 |

[a] Ref. 118.

**5.2. Vibrational Properties.** The simulated vibration spectrum represents a superposition of Infrared and Raman signals and shows a good match in wavenumbers with experimental data (Figure 4). The bands agree within 50 cm$^{-1}$ and include O-H stretching of hydroxide ions and OH groups at 3600 cm$^{-1}$, P-O stretching of phosphate at 1100 cm$^{-1}$ (strong) and 900 cm$^{-1}$ (weak), as well as O-P-O bending at ~600 cm$^{-1}$. Relative intensities and lattice modes below 400 cm$^{-1}$ are difficult to reproduce due to the lack of the full electronic structure in the force field. The experimental spectra also show broad signals for impurities such as water (from 3200 to 3500 cm$^{-1}$) and carbonates (near 1500 cm$^{-1}$) that do not appear in the simulation for pure hydroxyapatite.[123] The correlation between computed and experimental IR frequencies reflects the quality as well as limitations of the force field parameters.

**5.3. Mechanical Properties.** Mechanical properties are well reproduced in agreement with experiment (Table 6). There are five independent elastic constants for hydroxyapatite as a consequence of hexagonal symmetry. Isotropic Young's moduli $E_{iso}$, bulk moduli $K$, and isotropic shear moduli $G_{iso}$ of polycrystalline HAP samples were previously measured using ultrasonic interferometry and then converted into elastic constants using data on closely related fluoroapatite single crystals (Table 6).[137-139] Computed values with the force field deviate up to 20% and fall into a similar range as values reported by DFT calculations.[140-142] The 12-6 LJ potential yields slightly higher moduli and diagonal elastic constants in comparison to the 9-6 LJ potential. For example, $C_{11}$ and $C_{33}$ are 10% higher than experimental data using the 12-6 potential and 10% lower using the 9-6 potential. Somewhat higher stiffness using the 12-6 LJ potential results from the stronger repulsive part of the potential while atomic charges and bonded parameters are assumed to be the same for chemical consistency and simplicity. The anisotropic mechanical properties of HAP are well reproduced by 12-6 and 9-6 force fields as seen from the elastic constants $C_{11}$ and $C_{33}$.

Notably, the computed values $C_{44}$, $C_{12}$, and $C_{13}$ from PBE and GGA density functionals are less accurate in comparison to the force field, which performs very well considering the simplicity and a million times lower computational cost compared to DFT. A larger set of previous force field parameters using a Buckingham potential that was specifically parameterized to match elastic constants by de Leeuw et al. shows better match in elastic moduli but higher deviations of $C_{44}$, $C_{12}$, and $C_{13}$ compared to the force field reported here as well.[143] Therefore, the INTERFACE force field achieves very good agreement in elastic constants with experiment, in part even better than sophisticated quantum mechanical calculations.[140-142]

**Table 6.** Elastic moduli and constants in experiment and in simulation with the INTERFACE force field in 12-6 and 9-6 form (in GPa). Data from earlier simulations and density functional methods

are also shown. The force field achieves the same accuracy using one millionth of compute power in comparison to DFT.

| Method | Modulus | | | Elastic constants | | | | |
|---|---|---|---|---|---|---|---|---|
| | K | $E_{iso}^g$ | $G_{iso}^g$ | $C_{11}$ | $C_{33}$ | $C_{44}$ | $C_{12}$ | $C_{13}$ |
| Experiment[a] | 89.0 | 114 | 44.5 | 137 | 172 | 39.6 | 42.5 | 54.9 |
| Force Field 9-6[b] | 72.9 | 112 | 45.7 | 130 | 157 | 44.8 | 35.6 | 45.8 |
| Force Field 12-6[b] | 81.3 | 121 | 48.5 | 147 | 185 | 47.5 | 47.3 | 43.1 |
| Buckingham[c] | 90 | 118.8 | 46.6 | 134.4 | 184.7 | 51.4 | 48.9 | 68.5 |
| DFT/PAW-PBE[d] | 84.5 | 120.6 | 47.7 | 140.0 | 174.8 | 47.5 | 42.4 | 58.3 |
| DFT/PBE[e] | 90.7 | 106.6 | 46.6 | 145.2 | 191.4 | 37.6 | 47.8 | 73.6 |
| DFT/GGA[f] | 82 | 132.1 | 51.7 | 117.1 | 231.8 | 56.4 | 26.2 | 55.6 |

[a] Isotropic elastic moduli were measured on polycrystalline specimens of hydroxyapatite and single crystal fluoroapatite with an uncertainty of ±1% (K, $E_{iso}$, $G_{iso}$). Elastic constants for hydroxyapatite were then calculated using theory. See refs. [137-139]. [b] The uncertainty in computed elastic constants is ±1%. [c] Ref. [143]. [d] Ref [140] using DFT/PAW-PBE. [e] Ref. [141]. [f] Ref. [142]. [g] Using VRH approximation, ref. [144].

## 6. Cleavage Energy of Neat and Protonated Hydroxyapatite Surfaces in Vacuum

### 6.1. Cleavage Energy of Neat Hydroxyapatite Surfaces.
Similar to other minerals such as calcium silicates and clay minerals, surfaces of HAP reconstruct upon cleavage and do not recover the perfect original crystal structure upon agglomeration.[42, 130] Solid-vapor interface tensions are therefore not uniquely defined. A suitable measure for the surface energy of dry surfaces along various (h k l) planes is therefore the cleavage energy (Figure 5 and Table 6).[132] Molecular dynamics simulation was employed to monitor the cleavage of the common low index planes (001), (010), (020), and (101) of HAP in vacuum (Figure 5). (0 n 0) planes are thereby equivalent to (n 0 0) planes as a result of the hexagonal crystal symmetry. The cleavage energies upon slow, electroneutral cleavage of the neat phosphate-terminated surfaces range from 870 to 1220 mJ/m$^2$ according to the computation. The cleavage energy increases in the order (001) < (101) < (010) < (020) and consists of over 90% Coulomb energy and less than 10% van-der-Waals energy for all planes.[130]

In experiment, the freshly cleaved surfaces are hygroscopic, sensitive to air and moisture, and challenging to study (see section 4).[45-61] Direct measurements of cleavage energies in vacuum were thus not reported to our knowledge. Nevertheless, cleavage energies of minerals of comparable chemical composition were previously measured and the computed cleavage energies for hydroxyapatite are in good relative agreement (Table 6).[42, 73] Cleavage energies of calcium oxide and calcium hydroxide, for example, were measured by Brunauer to be 1300 and 1180 mJ/m$^2$ (±100 mJ/m$^2$).[120] Facet-averaged cleavage energies for HAP are expected slightly lower since the phosphate ions in $Ca_{10}(PO_4)_6(OH)_2$ are less polar than oxide or hydroxide ions (section 3).[91] The facet-averaged cleavage energy of HAP is dominated by the prismatic (010) and the basal (001) planes (Figure 2) and computed as 1000 ±50 mJ/m$^2$, in agreement with expectations of a cleavage energy slightly lower than that for calcium oxide and calcium hydroxide (Table 6). In contrast, computed cleavage energies using prior Buckingham potentials[111] and DFT calculations[18] are clearly overestimated at 1300 and 1700 mJ/m$^2$. Recent simulations with the INTERFACE force field also allow further comparisons with tricalcium silicate, $Ca_3SiO_5$, for which a cleavage energy of ~1300 mJ/m$^2$ was computed.[42] HAP has a lower Ca:P ratio in comparison to the Ca:Si ratio in Ca$_3$SiO$_5$ and is thus less ionic, and is also slightly hydrated so that a lower average cleavage energy of ~1000 mJ/m$^2$ is consistent with these trends. A further analogy is possible using the cleavage energy of the (001) plane of mica, which is reported as 375 mJ/m$^2$ in experiment.[112, 145] The mica basal plane contains K$^+$ ions at a similar area density as available Ca$^{2+}$ ions on the HAP (001) surface (2.1 versus 2.6 per nm$^2$). Cleavage energies are then expected to scale approximately with the square of the atomic charges of the superficial ions,[42] leading to a ratio $E_{Cleav,mica} / E_{cleav,HAP} \sim 0.4$ (q$^2$ (K$^+$) / q$^2$ (Ca$^{+2}$) ~ 1.0$^2$/1.5$^2$ ~ 0.4). The ratio of measured cleavage energy for mica and computed cleavage energy for HAP is ~0.43 and concurs with this analogy ($E_{cleav,mica}$ / $E_{cleav,HAP}$ ~ 375 mJ/m$^2$ / 870 mJ/m$^2$ ~ 0.43).[112]

Absolute values of cleavage energies are therefore highly consistent with experimental reference data. The calculations further show that the cleavage energy of hydroxyapatite notably depends on crystal facet (Figure 5 and Table 6). The (001) surface of HAP is of lowest cleavage energy with ~870 mJ/m$^2$ according to simulation. Cleavage of the (001) surface involves redistribution of calcium ions on the nascent surfaces, minor surface reconstruction, and comparatively minor local electric fields arise at the cleavage sites. Cleavage of the (101), (010) and (020) planes comprises the distribution of larger phosphate ions. The (101) surface is then of approximately 10% higher cleavage energy with ~980 mJ/m$^2$ according to the simulation, and reorientation of phosphate ions can be seen (Figure 5). The (010) and (020) surfaces exhibit the highest cleavage energy of ~1080 and ~1220 mJ/m$^2$ as phosphate ions with multiple negative charges are separated upon cleavage, accompanied by significant surface reconstruction and generation of noticeable local electric fields. For example, calcium ions on the (020) surface beneath the superficial layer of phosphate ions move to the outer surface upon cleavage to better shield the concentration of negative charge at the cleavage plane (Figure 5).

**6.2. Remaining Cleavage Energy of Protonated Hydroxyapatite Surfaces.** The surface energy of the protonated hydrogenphosphate and dihydrogenphosphate terminated surfaces at lower pH can be quantified by remaining cleavage energies in vacuum (Figure 5 and Table 6).[45-49] These energies are determined in a two-step process that involves first agglomeration of the equilibrated protonated surfaces followed by renewed cleavage and relaxation of the separated surface in molecular dynamics simulation. Perfect cleavage is then no longer possible and an uncertainty of ±5% is expected due to the overall series of modifications that involves (1) cleavage of neat HAP surfaces, (2) modification by protonation and relaxation of the surface, leading to an equilbrium protonated surface, (3) agglomeration of the protonated surfaces and interfacial relaxation, (4) and renewed separation of the agglomerated surfaces using molecular dynamics simulation to compute the remaining cleavage energies.[42]

The partial loss of calcium ions and the presence of less negatively charged protonated phosphate ions on the surfaces pH ~ 10 (Figure 1) reduce the remaining cleavage energy by 40% to 70% relative to the original cleavage energy of the bulk mineral, and by 70% to 80% relative to the original cleavage energy after two protonation steps at pH ~ 5. The remaining cleavage energies indicate that individual (h k l) cleavage planes respond rather differently to hydration and protonation relative to one another. The data also suggest notable differences in the relative stability of (h k l) facets when the pH conditions change.

In detail, the (001) surface shows a major decrease in cleavage energy from ~870 to ~250 mJ/m$^2$ after protonation to monohydrogen phosphate in the top layer. Further protonation to dihydrogen phosphate hardly affected cleavage further, arriving at ~220 mJ/m$^2$. The monohydroxylated (010) surface exhibits only a moderate decrease from ~1080 mJ/m$^2$ to 640 mJ/m$^2$ after formation of monohydrogen phosphate, related to removal of fewer calcium ions per surface area, a slightly higher negative charge on monohydrogen phosphate (Table 3), and uneven geometry (Figure 5). The second protonation step lowers the surface energy to ~320 mJ/m$^2$. The (020) surface also exhibits stepwise decreases in cleavage energy from ~1220 mJ/m$^2$ to ~400 mJ/m$^2$ and to ~250 mJ/m$^2$. The (101) surface showed stepwise decreases to ~390 and ~250 mJ/m$^2$ upon protonation to pH ~ 10 and pH ~ 5, respectively (Table 6). As a result, the (001) surface is of distinctly lower energy than the other surfaces at intermediate protonation to monohydrogen phosphate. At the stage of dihydroxylated surfaces, all four chosen surfaces are of similar cleavage energy within ±50 mJ/m$^2$ (Table 6). Similar compositions on every facet, less calcium ions per surface area and dominance of dihydrogen phosphate increase the structural similarity at pH ~ 5 (Figure 3). Also more degrees of freedom such as translation and rotation of dihydrogenphosphate ions are observed. It is also noted that, especially for the (010) surface, water penetration and protonation into more than the topmost molecular layer would likely decrease the reported cleavage energies further; therefore, the given values are a first guide.

Quantitative experimental data are scarce due to the difficulty to control surface protonation and hydration for a given number of subsurface layers. Aning et al.[146] examined slow cleavage of structurally similar fluoroapatite[118] in air and employed beam theory to calculate the cleavage energy in first approximation. The results were cleavage energies of 95 ±25 mJ/m$^2$ on (001) surfaces and 480 ±30 mJ/m$^2$ on (010) surfaces. The result for the (010) plane is in the same range as computed values for partly protonated surfaces (Table 6). On the other hand, large differences between the computed and measured cleavage energies for the basal (001) plane may be associated with difficulties in humidity control during the measurement.[146, 147] Further experimental data for HAP surfaces also include solid-vapor surface tensions by contact angle measurements on the basal plane (001) and on the prismatic plane ((010), (020)) by Busscher and van Pelt.[148, 149] However, surface tensions refer to surfaces after cleavage in contact with water and are barely related to cleavage energies.[130, 132] The measured values of 89 ±24 mJ/m$^2$ and 80 ±9 mJ/m$^2$ for HAP basal and prismatic planes, respectively,[148, 149] nevertheless suggest similar interface energies of the different HAP crystal facets consistent with the similarity of the computed cleavage energies of these facets at pH 5.

More evidence for the computed facet-averaged surface energies of monohydrogen phosphate and dihydrogen phosphate terminated HAP can be derived by comparison to available measurements for C-S-H (Ca$_3$Si$_2$O$_7 \cdot$ 2 H$_2$O) and selenite (CaSO$_4 \cdot$ 2 H$_2$O) of 386 and 352 mJ/m$^2$.[150, 151] These compounds reflect left (Si) and right (S) neighbors to Ca phosphates (P) in the periodic table, are of similar polarity, and contain some crystal water to reflect initial hydration or protonation, respectively. The surface energies in the range 300 to 400 mJ/m$^2$ are similar to the surface energy of hydroxyapatite at pH ~7 (between monohydrogen phosphate and dihydrogen phosphate termination). Future studies may also reveal more details of the impact of water penetration and protonation deeper than 0.5 nm on cleavage and cohesive properties.

**Table 6.** Computed cleavage energy of hydroxyapatite facets in neat and hydrated form in comparison to available experimental data in vacuum. Data are given in mJ/m$^2$. The uncertainty of computed values is ±25 mJ/m$^2$, supported by the difference between 12-6 and 9-6 force fields.

| Method | Cleavage energy | | | |
|---|---|---|---|---|
| | Basal plane (001) | Prismatic plane (010) | (020) | (101) |
| **1. HAP surface as-is (PO$_4^{3-}$ terminated)** | | | | |
| Force field 9-6 | 880 | 1076 | 1234 | 984 |
| Force field 12-6 | 869 | 1092 | 1213 | 973 |
| Buckingham potential[a] | ~1100 | ~1310 | | ~1048 |
| DFT-B3LYP[b] | 1058 | 1709 | NA | 1646 |
| *Related experimental data* | | | | |
| HAP | NA | | | |
| CaO (more ionic)[c] | 1310 ± 200 (facet average) | | | |
| Ca(OH)$_2$ (more ionic)[c] | 1180 ± 100 (facet average) | | | |
| **2. HAP surfaces, HPO$_4^{2-}$ terminated (pH ~ 10)** | | | | |
| Force field 9-6 | 250 | 658 | 411 | 369 |
| Force field 12-6 | 243 | 618 | 386 | 408 |
| **3. HAP surfaces, H$_2$PO$_4^-$ terminated (pH ~ 5)** | | | | |
| Force Field 9-6 | 225 | 330 | 250 | 271 |
| Force Field 12-6 | 222 | 312 | 262 | 221 |
| *Related experimental data (similar compounds)* | | | | |
| Fluoroapatite[d] | 95 ± 25 | 480 ± 30 ((010) and (020)) | | NA |
| CaSO$_4$ · 2 H$_2$O[e] | 352 ± 20 (facet average) | | | |
| Ca$_3$Si$_2$O$_7$ · 2 H$_2$O[f] | 386 ± 20 (facet average) | | | |

[a] Buckingham potential, Ref. 111. [b] Quantum mechanical calculations, Ref. 18. [c] Ref. 120. [d] Ref. 146. Slow cleavage of fluoroapatite (FAP) was performed at the basal and prismatic planes. Simultaneous hydration is likely to play a role and the data are therefore somewhat uncertain. The

cleavage energy for the prismatic plane may include contributions by (010) and (020) facets. [e] Gypsum, ref. [151]. [f] Tobermorite, ref. [150].

# 7. Interaction of Hydroxyapatite Surfaces with Water and the Structure of Solid-Liquid Interfaces

The enthalpy of immersion and structure of the solid-liquid interfaces was examined for the same Miller planes (Figure 6 and Table 7). Freshly cleaved, chemically unmodified hydroxyapatite surfaces maintain a distinct boundary with the liquid phase (Figure 6, left column) which becomes increasingly disordered and diffuse upon protonation (Figure 6, center and right columns). The quantitative representation of pH values of the surface shows that also dissolution processes can be visualized in a simulation for the first time. As the pH value decreases, the hydration energy is reduced in a similar way for all facets. Reduced hydration energies at lower pH values correlate with leaching of hydroxide ions, reduced area density of calcium ions on the surfaces, and decreased atomic charges on the oxygen atoms of monohydrogenphosphate and dihydrogenphosphate ions in comparison to phosphate ions (Figure 1 and Table 3).

**7.1. Immersion Energy.** Immersion energies for the different facets provide a quantitative measure for the affinity of hydroxyapatite towards water and enable comparisons to experimental data from calorimetry and adsorption isotherms (Table 7).[60, 152, 153] Computed values of the energy of immersion range from 520 to 800 mJ/m$^2$ for the surfaces at pH 10 and from 360 to 620 mJ/m$^2$ for the same surfaces at pH 5 depending on facet. Values at the high end are seen for (010) facets, related to hydration and protonation of only the topmost phosphate layer on this corrugated surface. The characteristic roughness of the (010) surface can be relieved by reconstruction to a (020) surface upon hydration and partial dissolution. Typical values for protonation of the entire exposed surface are in the range 500 to 670 mJ/m$^2$ at pH 10 and 350 to 500 mJ/m$^2$ at pH 5, respectively. The energy of immersion would likely decrease also further when initial hydration and protonation extend beyond the topmost molecular layer, as assumed in this study, to include additional subsurface layers.

**7.2. Comparison to Measurements.** Available measurements of immersion energies and immersion free energies range from 234 to 700 mJ/m$^2$, generally in the same range (Table 7).[60, 152, 153] The experiments were performed on polycrystalline (pellet-pressed) substrates near neutral pH. Unfortunately, polycrystalline samples have no defined facet composition and sample preparation differed from one study to another. Several samples were prepared by precipitation from hot solution and resulted in low immersion free energies (234 to 432 mJ/m$^2$).[152] The surfaces in these studies were already hydrated before the measurement, related to the high affinity of Ca$^{2+}$ ions to water, and reported immersion energies of HAP are too low. One set of samples was prepared by solid state reaction at 1200 ºC and resulted in an immersion free energy up to 476 mJ/m$^2$.[152] Even in this case, superficial hydration in air prior to measurement was still possible. The immersion energy of perfectly dry HAP crystals was determined to be 600 to 700 mJ/m$^2$ by calorimetry as a function of increasing temperature before dehydration and introduction of P-O-P bonds occurred.[60, 153] The calorimetric data appear most justified and compare well with simulation results. To compute a facet-average immersion energy from MD simulation, the immersion energies for various (h k l) facets can be weighted according to their contribution to the overall nanocrystal shape (Figure 2). Data by Tanaka[60] also specify the ratio of HPO$_4^{2-}$ and H$_2$PO$_4^-$ on HAP surfaces as 31:69 at neutral pH near 7. If equal weight of the four facets is assumed, the computed facet-average immersion energy at pH 7 is 629 × 0.31 + 490 × 0.69 = 533 mJ/m$^2$ (Table 7). More realistically, data from crystal growth and computed immersion energies indicate that the (010) facet is dominant (Figure 2). The computed facet-average immersion energy at neutral pH then amounts to 619 mJ/m$^2$, assuming a contribution of the (010) facets of 70% and of each remaining facet of 10% each. This shape-average immersion energy correlates well with the calorimetry measurements int eh range of 600-700 mJ/m².

The experimental data for comparison include calorimetrically measured energies of hydration, $\Delta E_{hyd}(=-\Delta E_{imm})$ and free energies of hydration obtained from adsorption isotherms, $\Delta A_{hyd} = \Delta E_{hyd} - T\Delta S_{hyd}$, which include changes in entropy upon adsorption of water. The mobility of water is reduced on all surfaces upon adorption, especially on the monohydrogen-phosphate terminated (010) surface (Figure 6). A maximum entropy-related contribution of $-T\Delta S_{hyd}$ ~ +90 mJ/m² can be derived from the entropy of freezing of water $\Delta S$ = -22 J/ (mol · K)[61] and an area density of about 100 molecules per (3.5 nm)² surface area. The true difference is only estimated as +30 to +50 mJ/m² since adsorbed water molecules retain significant rotational and translational mobility in comparison to crystalline ice. In effect, the negative energy of hydration is reduced to a less negative free energy of hydration. In other words, the positive energy of immersion is 30 to 50 mJ/m² higher than the free energy of immersion (note that the energy and free energy of immersion are conventionally stated with a positive sign even though immersion is an exothermic process). Accordingly, a free energy of immersion of 476 mJ/m² based on adsorption isotherms (Table 7) corresponds to an energy of immersion of ~516 mJ/m². The overall correlation between simulation (~620 mJ/m²) and experiment (520 to 700 mJ/m²) for the facet averaged energies of immersion is very good.

**7.3. Facet-Specific Trends.** A benefit of the computational models is insight into hydration properties as a function of pH and facet-specific detail, which has not been accessible by experiment. The results show that the (010) surface is the most attractive surface for water in a pH range from 10 to 5, corresponding to an immersion energy of ~800 to ~620 mJ/m², respectively. The high affinity of the prismatic (010) plane to water correlates with its dominance of the surface area of hexagonal, rod-like HAP crystals.[154-156] The surface with the least affinity to water is the

(001) surface, also across a pH range from 10 to 5, corresponding to an immersion energy of ~520 to ~360 mJ/m², respectively. In agreement, the (001) basal plane is known to cover less surface area and is the fastest-growing plane on hexagonal HAP crystallites.[154-156] These facet-specific differences suggest correlations between the affinity towards water and preferred nanocrystal shape.

Hydration energies of apatite surfaces therefore depend on pH, crystal facet, as well as on the initial depth of water penetration and protonation. Further detailed studies using the atomistic models may elucidate the relative stability of facets and apatite nanocrystals of different shape under various conditions.

The relative stability of facets could also be evaluated by solid-liquid interfacial tensions ($\gamma_{SL} = \Delta A_{Hydr} + \gamma_{SV} + \gamma_{LV}$). The difficulty for reconstructing and dynamic surfaces is, however, that the surface free energy $\gamma_{SV} \sim E_{Cleav}$, the hydration energy $\Delta A_{Hydr} \sim -E_{Imm}$, and the interfacial tension $\gamma_{SL}$ become a function of the initial depth of water penetration and protonation in addition to pH and (h k l) crystal facet. Approximate estimates of $\gamma_{SL}$ using available values $\Delta A_{Hydr} \sim -E_{Imm}$, $\gamma_{SV} \sim E_{Cleav}$, and the water surface tension ($\gamma_{LV} = 72$ mJ/m²) for monolayer protonation from Figures 5 and 6 are slightly negative, suggesting a trend toward dissolution and further hydration of subsurface layers.

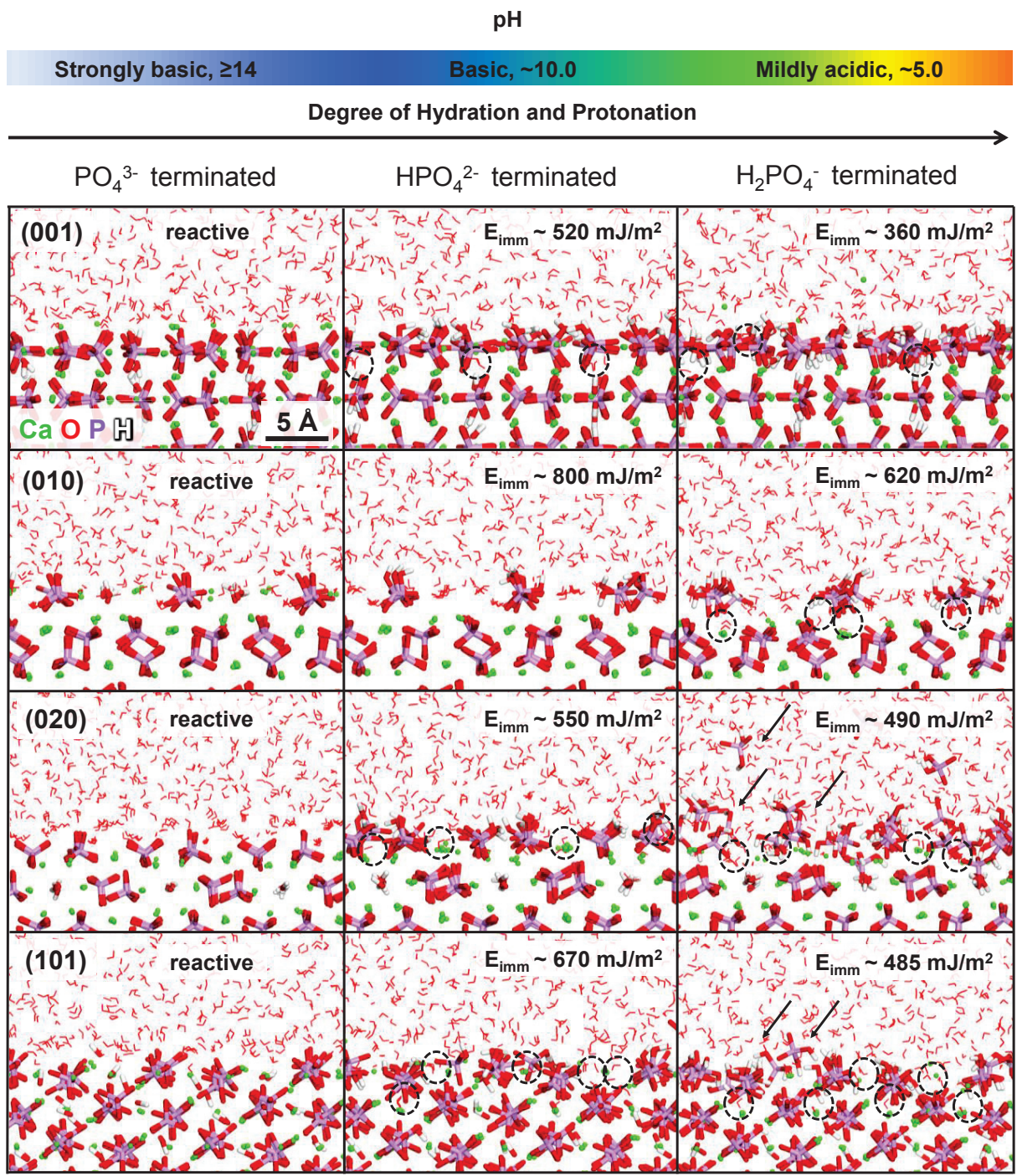

**Figure 6.** Snapshots of hydroxyapatite-water interfaces at different pH and computed energy of immersion (= energy of hydration without negative sign). Freshly cleaved, $PO_4^{3-}$ terminated surfaces are reactive and only stable at very high pH (left column). Surfaces terminated by $HPO_4^{2-}$

ions at pH ~ 10 exhibit significant, facet-specific immersion energies (middle column). Dashed black circles highlight penetration of water molecules into the surface of the crystal lattice. Surfaces terminated by $H_2PO_4^-$ ions at pH ~ 5 exhibit lower immersion energies (right column). Protonated phosphate ions on the surface are more mobile, assume random orientations, and partially dissolve from the surface, especially from (020) and (101) surfaces, as observed also in experiment (indicated by black arrows).

**Table 7.** Energy of immersion of hydroxyapatite surfaces according to simulation and experimental data (in $mJ/m^2$). Energy expressions with a 12-6 LJ potential yield 2-10% higher immersion energies than the energy expressions with a 9-6 LJ potential.

| Method | Immersion energy ($mJ/m^2$) | | | |
|---|---|---|---|---|
|  | (001) | (010) | (020) | (101) |
| 1. Computed at pH ~ 10, $HPO_4^-$ terminated | | | | |
| 9-6 LJ[a] | 509 | 753 | 518 | 649 |
| 12-6 LJ[a] | 530 | 820 | 572 | 683 |
| Facet average (equal weight; real weight)[d] | 629; 734 | | | |
| 2. Computed at pH ~ 5, $H_2PO_4^-$ terminated | | | | |
| 9-6 LJ[a] | 354 | 612 | 476[b] | 481[b] |
| 12-6 LJ[a] | 367 | 631 | 509[b] | 492[b] |
| Facet average (equal weight; real weight)[d] | 490; 567 | | | |
| Experimental data (pH ~ 7) | | | | |
| Adsorption isotherm of water (precipitated and dry samples)[e] | 365.3, 432.5, 476.3 (20°C) 234.5, 373.5, 407.3 (25°C) | | | |

| | |
|---|---|
| Calorimetry (dry/hot samples)[f] | ~600 (25°C) |
| Calorimetry (dry/hot samples)[g] | ~700 (25°C) |

[a] The uncertainty of simulated immersion energy is ±2%, or ±10 to ±15 mJ/m². [b] Some dihydrogen phosphate ions begin to dissolve during the simulation. [d] An equal contribution of 25% of each facet to the total immersion energy is assumed, as well as a more practical formula in which the dominant (010) surface makes 70% contribution and 10% contributions by the remaining facets. [e] Ref. 152. Note that free energies of immersion are reported here. The corresponding energies of immersion can be obtained by addition of +40 mJ/m², i.e., up to 516 mJ/m² (see text). [f] Ref. 60. [g] Ref. 153.

**7.4. Density Profiles of Phosphate Ions, Calcium Ions, and Water.** Density profiles of the hydroxyapatite-water interfaces reveal specific details for each facet at different pH values (Figure 7). At first sight, the density profiles indicate unique facet-specific spacing of phosphate and calcium ions. The variation in sequence and distance of these ions at the surface is a major reason for the differential adsorption of water (Table 7), as well as for the differential adsorption of biological and organic molecules. The 2$^{nd}$ or 3$^{rd}$ sub-surface average atomic positions of phosphorus were chosen as a reference coordinate of 0 Å in Figure 7. In the present models, hydration and protonation was restricted to the topmost layer of phosphate ions, i.e., about 5 Å depth.

The density profiles visualize the loss of calcium at the surface with decreasing pH for all four facets, leading to a lower Ca:P ratio as observed in NMR measurements.[52-55] This is a result of the progressive hydration and dissolution of calcium hydroxide from pH>14 towards pH~5, which leads to dissolution of much of the Ca$^{2+}$ content at the aqueous interfaces (Figure 1). In the same sequence of pH, the definition of the outer phosphate atomic positions near the solid-liquid interface is also reduced on all facets, as seen by decreasing local maxima in phosphate density towards lower pH (Figure 7). The broadening of superficial phosphate positions is associated with a randomization of its orientation on the surface. At the aqueous interface of (020) and (101) facets at pH values of 10 to 5, partial dissolution of $HPO_4^{2-}$ ions and $H_2PO_4^-$ ions is also observed while (001) and (010) facets exhibit more resistance to phosphate dissociation. The observations are consistent with experimental observations of higher solubility of dihydrogen phosphate in water compared to monohydrogen phosphate.[61] Thereby, the observation of ion dissolution at the surface does not at all affect the structural stability of the neat mineral a few atomic layers below as verified by simulations over more than 100 ns duration.

The density profiles of superficial water exhibit visible layering at higher pH, related to higher surface charge and cation density of apatite (Figure 7). Close to the outer layer of calcium ions and phosphate ions on the hydroxyapatite surface, two to four recognizable water layers are observed that converge to the bulk density of water at about 5 to 10 Å distance. The first layer sometimes features two separate peaks, for example, on (001) and (101) surfaces, that are related to the surface corrugation of alternating phosphate and calcium ions by about 1 Å. Lower pH values reduce the definition of distinct water layers, except on the strongly corrugated (010) surface (Figure 6). The observations of interfacial layering of water are similar to those reported for other oxide surfaces.[67] Moreover, water penetration up to about 5 Å into the subsurface layers of $Ca^{2+}$ and phosphate is observed on all facets at lower pH values (see highlights in Figure 6 and density profiles in Figure 7). Lower surface charge and accompanying surface disorder at lower pH smoothen the water density profile and ease the penetration of water into the mineral.

It appears likely that water penetration as well as partial dissolution of hydrogenphosphate ions would extend deeper into the surface when more surface layers participate in the hydration reaction. The level of penetration may be controlled by moisture level, temperature, and exposure time in experiment, and can be represented by the depth of $Ca(OH)_2$ leaching and phosphate protonation in the models. Details remain to be explored in future studies by appropriately modified atomistic models and experimental measurements.

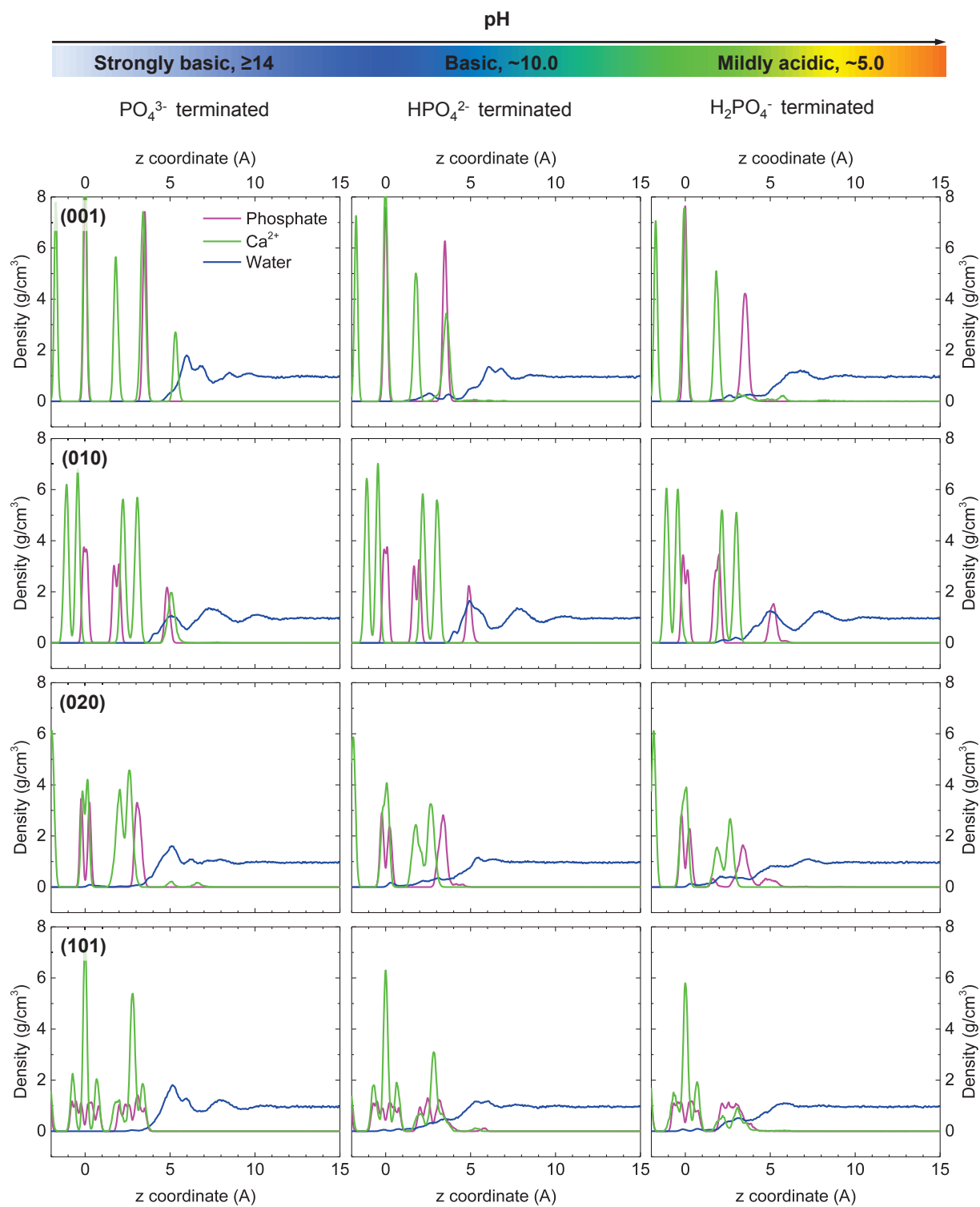

**Figure 7.** Density profiles of phosphate ions, calcium ions and water at the (001), (010), (020), and (101) hydroxyapatite surfaces at different pH values, assuming hydration of the top molecular

layer on the surface. The density profile of phosphate including all protonated phosphate species at the surface, calcium, and water differs from facet to facet. The loss of superficial calcium ions can be seen on all facets upon decreasing pH value. Less reconstruction is observed on phosphate-terminated surfaces at high pH values in comparison to monohydrogen phosphate-terminated surfaces and dihydrogen phosphate-terminated surfaces at low pH values, particularly on the (020) and the (101) planes. The water density profiles indicate layering and short-range order near the surfaces. At low pH, water density fluctuations near the surface are reduced and penetration into subsurface layers occurs.

**7.5. Biomolecular Adsorption.** The adsorption of peptides, DNA, and other polymers to various apatite surfaces can be studied in aqueous solution using the proposed force fields (e.g. CHARMM-INTERFACE) and surface models, just as previously described for silica and other minerals.[41, 42, 112, 114] Examples as a proof of concept are shown for the conformation of a HAP-binding peptide identified by phage display, S(+)VSVGGK(+)(-)·Cl(-),[22, 65] on the (001) basal plane of the mineral as a function of pH (Figure 8). At pH 10, negatively charged carboxylate groups in the peptide interact with calcium ions on the surface, and ammonium groups at the N-terminal end are in close contact with hydrogenphosphate ions on the surface. Ion pairing and ion exchange are dominant contributions to adsorption. At pH 5, different residues such as serine (S1) and valine (V2) are attracted to the apatite surface, related to the lower area density of cations (Figure 8). Binding is then mostly mediated by hydrogen bonds and hydrophobic interactions.[40, 41] A detailed discussion of selective adsorption as a function of biopolymer sequence, facet, and pH will be reported separately (see preview of data in refs. [66],[39]).

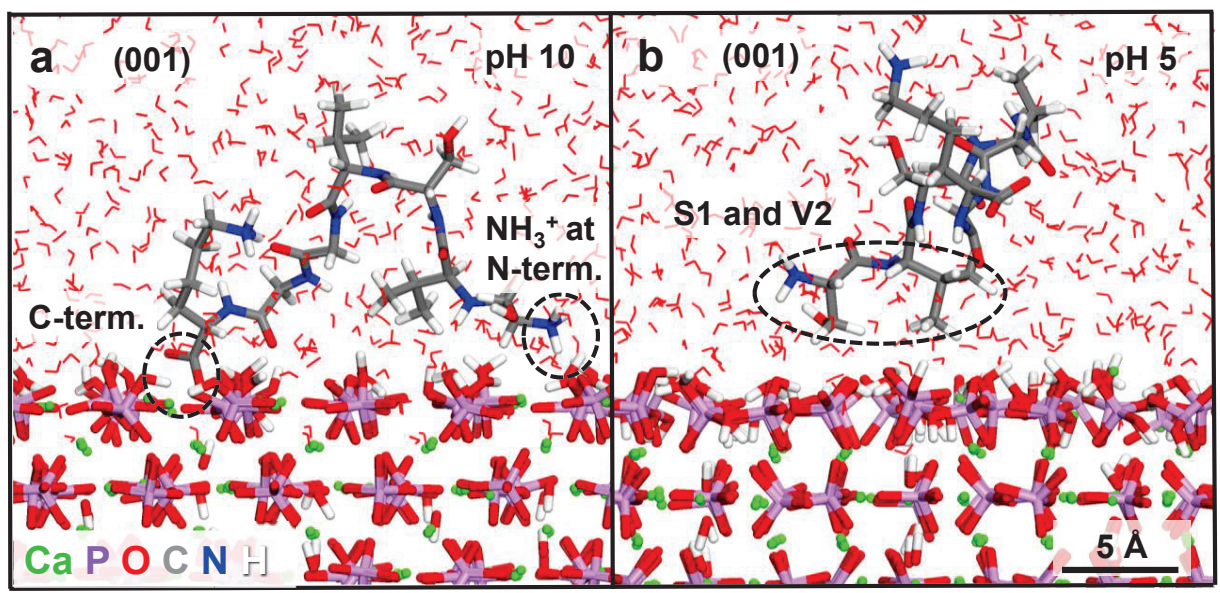

**Figure 8.** Adsorption of peptide SVSVGGK on the hydroxyapatite basal plane (001). Representative conformations and binding groups change significantly for pH values of 10 versus 5. Ionic groups are attracted to the surface at pH 10 while hydrophobic/polar groups are attracted to the surface at pH 5.

## 8. Conclusion

Force field parameters and surface models of hydroxyapatite were introduced and extensively tested for a range of different facets and pH values in aqueous environment. The force field parameters are compatible with numerous harmonic energy expressions, including PCFF, COMPASS, CHARMM, DREIDING, GROMACS, AMBER, CVFF, and OPLS-AA. The parameters are also part of the INTERFACE force field with accurate parameters for other minerals and metals.[73] The new parameters and models facilitate simulations of aqueous, organic and biomolecular interfaces of apatites under realistic solution conditions including pH and ionic strength, which is a large improvement over earlier models. The force field for hydroxyapatite is developed in the context of the INTERFACE force field as a uniform simulation platform for all types of materials, including thermodynamic consistency for each compound and standard combination rules of non-bond parameters. It reproduces atomic scale as well as bulk properties in excellent agreement with laboratory measurements, including atomic charges, van-der-Waals radii, bonded parameters, lattice parameters, cleavage energies, hydration energies, and elastic constants. Computed lattice parameters deviate less than 0.5% from XRD data, the cleavage energy of HAP surfaces agrees with that of chemically similar compounds, computed elastic constants are of the same accuracy as results from quantum mechanical calculations (deviation of ~10% from measurements), and the computed IR spectrum approximates experimentally measured frequencies of O-P-O bending, P-O stretching, and O-H stretching vibrations well.

A suite of atomistic models for hydrated and protonated hydroxyapatite surfaces are introduced for (h k l) facets that cover a range of pH values using abundant experimental data on acid-base equilibria and surface spectroscopy as a foundation. Hydration energies are found to be facet-specific and diminish toward lower pH, accompanied by an increase in surface disorder and partial dissolution. These trends correlate with the area density of ionic groups on the (h k l) facets, and computed hydration energies are in good agreement with facet-average experimental measurements at pH ~ 7. The models for apatite surfaces can be adjusted to different degrees of hydration, protonation, and surface penetration. Defects such as carbonate, magnesium, and fluoride ions can also be incorporated exploiting the coverage of the INTERFACE force field, including INTERFACE-PCFF, INTERFACE-CHARMM, and other energy expressions. First examples of highly specific binding of peptides demonstrate the applicability to biological interfaces that will be explained in separate contributions in detail.

The force field and the surface model database provide a new computational tool to explore mechanisms of specific recognition and assembly of apatite-biological interfaces on the 1 to 1000 nm scale. Quantitative insight in atomic resolution can be obtained in comparison with laboratory and clinical studies using molecular dynamics and Monte Carlo methods. Applications are envisioned to understand the structure of bone and teeth, osteoporosis and bone implants, caries and dental implants, protein-driven apatite mineralization and demineralization, drug screening on apatite surfaces, as well as calcification in arteries. Moreover, biomechanics of protein-apatite composites can be studied n high accuracy at length and time scales a million times larger than those accessible by quantum mechanical methods. In summary, realistic models and simulations of apatites have the potential to contribute towards solutions of grand medical challenges in combination with experiment and clinical tests. The aim of this work is to provide sufficient detail and initial validation of apatite models to open up and advance such exciting research opportunities.

**Acknowledgements**

The authors acknowledge support by the National Science Foundation (DMR 0955071, 1437355), UES, Inc., the Air Force Research Laboratory, Procter and Gamble Co, the Office of Naval Research (ONR-MURI N00014-14-1-0675), the University of Akron, and the University of Colorado-Boulder. The allocation of computational resources at the Ohio Supercomputing Center is also acknowledged.

**Associated Content**

The Supporting information contains further details of the surface models and a complete description of the computational methods. The hydroxyapatite surface models are contained in the latest release of the INTERFACE force field (v1.5) and attached as a separate file including documentation.

**Supporting Information**

For

**Accurate Force Field Parameters and pH Resolved Surface Models for Hydroxyapatite to Understand Structure, Mechanics, Hydration, and Biological Interfaces**

by

Tzu-Jen Lin[1] and Hendrik Heinz[1,2*]

Department of Polymer Engineering, University of Akron, Akron, OH 44325, USA

Department of Chemical and Biological Engineering, University of Colorado-Boulder, Boulder, CO 80309, USA

* Corresponding author: hendrik.heinz@colorado.edu

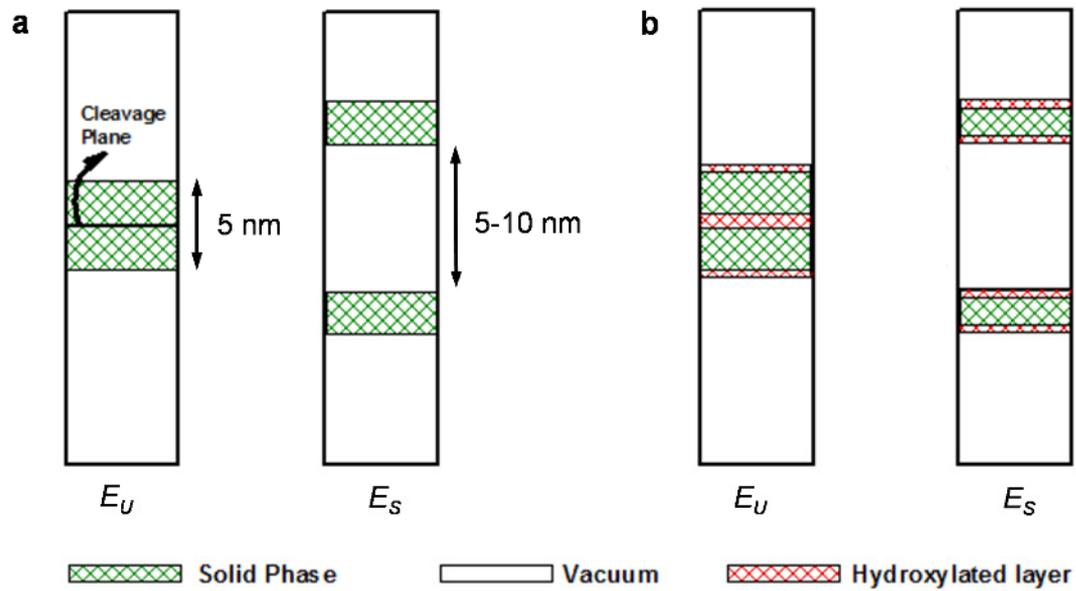

**Figure S1.** Schematic illustration of cleavage (a) of the neat hydroxyapatite surface and (b) of the hydrated and protonated hydroxyapatite surface. The cleavage energy was computed as a difference in average energies $E_{Cleav} = E_S - E_U$.

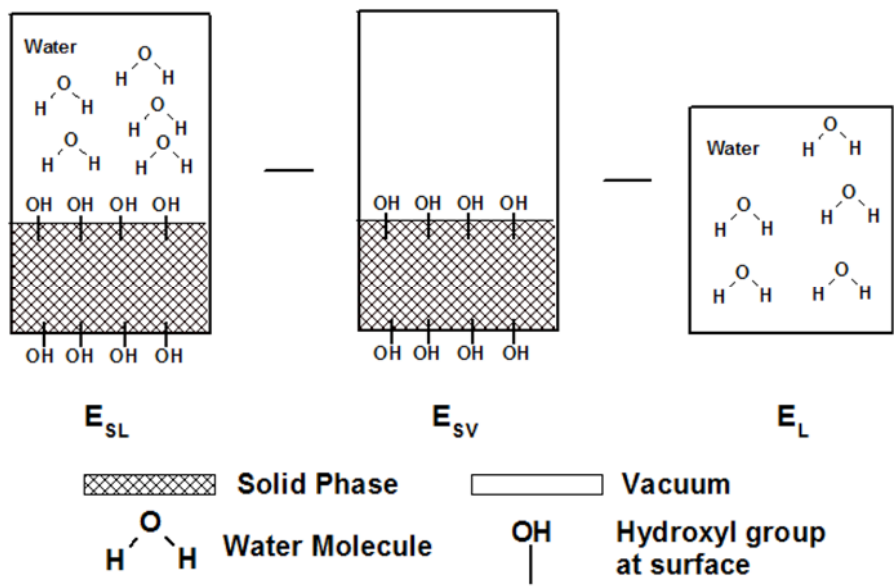

**Figure S2.** Illustration of the calculation of the energy of hydration and the energy of immersion using three simulation boxes (projection in 2D). The exothermic energy of hydration $\Delta E_{Hydr}$ is calculated as $\Delta E_{Hydr} = (E_{SL} - E_{SV} - E_L)/(2A)$, whereby $A$ is the cross-sectional area of the box. The energy of immersion is $E_{Im} = -\Delta E_{Hydr}$, i.e., the same value with positive sign.

## S1. Further Details of Hydroxyapatite Surface Structure and Molecular Models

The chosen surface models are informed by convergent laboratory measurements and knowledge on acid-base chemistry as summarized in the main text.[1-26] The data include X-ray diffraction, measurements of acid-base equilibria, solubility products, adsorption isotherms of water, AFM, SEM, TEM, IR and NMR spectroscopy, XPS, and Auger electron spectroscopy. Implementation of this long known information into molecular models might be the most important contribution in this work. It lays a foundation for the application of powerful simulation techniques to realistic apatite models under given solution conditions and disruptively improves the level of accuracy from prior models that neglected surface chemistry and pH. The known data explain how the surface of hydroxyapatite differs in composition from core HAP upon contact with water, as a function of pH, temperature, and other conditions, providing a wealth of time-proven and reproducible input for molecular models.

Rootare et al.[26] showed that water vapor can chemically adsorb on the hydroxyapatite surface with an area density of 8.7 water molecules per nm$^2$. This density corresponds to a molecular monolayer and is about twice as much as the phosphate area density of about 4 groups per nm$^2$ (Table 3), thus enabling up to two protonation reactions leading to dihydrogen phosphate. Rootare[10] also studied the solubility of HAP and proposed surface reactions that lead to the dissolution of hydroxide and formation of superficial monohydrogen phosphate ions as implemented in the models. The initial steps are:

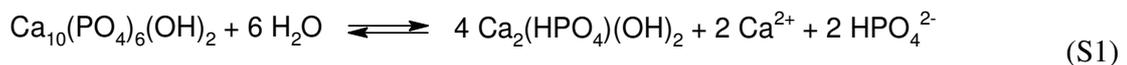
$$Ca_{10}(PO_4)_6(OH)_2 + 6\ H_2O \rightleftharpoons 4\ Ca_2(HPO_4)(OH)_2 + 2\ Ca^{2+} + 2\ HPO_4^{2-} \quad (S1)$$

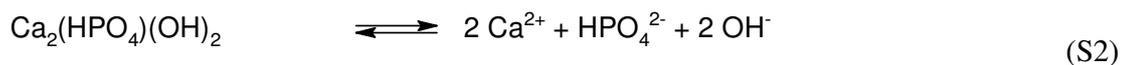
$$Ca_2(HPO_4)(OH)_2 \rightleftharpoons 2\ Ca^{2+} + HPO_4^{2-} + 2\ OH^- \quad (S2)$$

The reactions are anticipated to occur on the surface whereby the complex resulting from reaction (S1) covers the surface and strongly bonds to bulk HAP underneath. The authors also examined the hydrolysis and solubility at different concentration and ionic strength. The proposed atomistic models represent apatite surfaces in agreement with these suggestions and further distinguish crystallographic facets (Figure 5). The reactions described in equations (3) to (6) in the main text are specific to these surface environments and distinguish surface-bound groups from dissolved groups.

Studies by another group involving Bertazzo et al.[12] have shown that proton adsorption on the surface of calcium phosphate near physiological pH is mainly due to the presence of $HPO_4^{-2}$ ions. Measurements of the dissolution of HAP surfaces and related calcium phosphates of adjusted stoichiometry were employed to determine the ionic product of HAP for pH values between 5.0 and 8.0.[11] The proposed mechanisms of protonation and dissolution by Bertazzo et al. are consensual with those by Rootare and Craig[10] and stress the significant solubility products of $CaHPO_4$ (pK=6-8) and of $Ca(H_2PO_4)_2$ (pK=6-13) as a function of pH. These facts are also consistent with the ability to observe dissolution on the HAP surface using the new molecular models in the simulation (Figures 6 and 7). A similar view to dissolution was taken by Dorozhkin.[22,23] Dorozhkin considers the hydroxide neutralization by an acid as a logical first step due to high pK of water in comparison to $pK_3$ of phosphoric acid. Then, protonation to monohydrogen phosphate ensues towards lower pH, supported by measurements of superficial Ca:P ratios close to 1.0, and further formation of dihydrogen phosphate. Dorozhkin's explanations are well supported by chemical knowledge and experimental validation.

Jager[13] also found that nanocrystalline hydroxyapatite prepared in aqueous environment has different composition on the surface in comparison to that of the core using solid-state NMR spectroscopy. The nanocrystals consist of a crystalline core HAP and of a disordered surface region. The disordered surface region was described to be of a thickness of about 1 nm that is dominated by monohydrogen phosphate anions with no HAP-like structural motif. Jager suggested that the surface composition is close to octacalcium phosphate (OCP), $Ca_8(HPO_4)_2 (PO_4)_4 \cdot 5 H_2O$. Octacalcium phosphate is a crystalline phase of calcium phosphate with an alternative layered structure in which one layer contains pure phosphate as in HAP, and the other contains water, calcium and hydrogen phosphate ions.[14, 15]

The composition of the protonated surface layer of HAP appears not well defined from an experimental viewpoint. The range of composition may be expressed as $Ca_{(10-X)}(HPO_4)_X(PO_4)_{(6-X)}OH_{(2-X)}$,[16] whereby the limiting composition of HAP is somewhat uncertain. Brown and Martin[17] suggested the x value is 1 with Ca/P =1.5; Berry[18] suggested the x value is 2 with Ca/P =1.33; Meyer and Fowler[19] claimed the existence of a limiting value of Ca/P = 1.4 which indicates that the surface would consist of a layer with the composition $Ca_{8.4}(HPO_4)_{1.6}(PO_4)_{4.4}(OH)_{0.4}$. From a viewpoint of molecular models, aggregate formulas depend on how many molecular layers on the surface are taken into consideration, as inclusion of additional neat phosphate/hydroxyapatite layers deeper underneath the surface continually changes the stoichiometry towards higher phosphate and hydroxide content. In addition, there are differences in layer-by-layer composition according to (h k l) facet. Therefore, all of the above compositions could be implemented in the proposed atomistic models, depending on how many surface layers are hydrated and protonated, and how many layers are taken into consideration when formulating an aggregate composition. The atomistic models fully capture possible variations, including the effect of pH and (h k l) facet, as well as the number of hydrated layers with altered stoichiometry (Table 3).

Robinson[20] studied enamel apatite surfaces by AFM. The AFM study was carried out using carboxylated cantilever tips and mature crystals between pH 2 and 10. The adhesion force between the tip and HAP surface increased from pH 10 to a maximum at pH 6.6 because of increased hydrogen bonding resulting from protonation of the phosphate at the surface. The adhesion force dropped dramatically and became variable below pH 6.6 due to removal of fully protonated phosphate from the surface. IR studies also detected the formation of protonated surface phosphate sites in apatites. Ishikawa assigned vibration in the range 3650-3680 cm$^{-1}$ to OH stretching in surface HPO$_4$ and H$_2$PO$_4$ groups.[21] A more recent infrared spectroscopic study has shown that the apatite surface consists of HPO$_4^{2-}$ and H$_2$PO$_4^-$ ions in a ratio of 31:69 at pH = 7.[24] In addition, very well defined values of acid-base constants for phosphoric acid are available, including pK$_1$ = 2.15, pK$_2$ = 7.20, and pK$_3$ = 12.67 at 25 °C.[25] The associated ranges of stability of PO$_4^{3-}$ ions, HPO$_4^{2-}$ ions, H$_2$PO$_4^-$ ions, and H$_3$PO$_4$ molecules have been matched to fully agree with the reported surface compositions above as a function of pH.

Therefore, data from multiple sources since the 1960's have been available to characterize the surface structure of hydroxyapatite. pK values of H$_3$PO$_4$, infrared spectroscopy, adsorption data, the analysis of Ca:P ratios near the surface, chemical expectations of acid base equilibria at the surface, as well as mechanistic suggestions of hydration and protonation provide convergent information. The data were carefully analyzed to introduce the new surface models of hydroxyapatite in full consistency. In this work, we only modified the top surface layer and subsurface layer protonation up to several nm into the bulk crystal can easily be implemented following the same protocol in future work. The surface models are contained in the latest release of the INTERFACE force field (v1.5) and attached as a separate file including documentation.

## S2. Computational Details and Methods

**S2.1. Rationale of Force Field Derivation and Validation.** The derivation and validation of force field parameters and surface models follows earlier established protocols as described in the INTERFACE force field in general, and specifically for clay minerals,[27] fcc metals,[28] pH dependent surfaces of silica,[29, 30] tricalcium silicate, and tricalcium aluminate (Figure 3).[31, 32] Computational methods for parameters testing involve molecular mechanics and molecular dynamics simulations to (1) calculate density, surface (cleavage) energy, elastic modulus, IR spectrum, hydration energies and density profiles, (2) refine the force field parameters, and re-enter this loop with an updated chemical rationale until consistently low deviations are achieved (Figure 3). Initial tests of the parameters focus on density, lattice parameters, cleavage energy, and IR spectra, including the exploration of nearby ranges of atomic charges. Lattice parameters, cleavage energies, hydration energies, and elastic moduli usually result in good agreement with experiment after the initial iterations and were used for secondary validation and parameter refinements.

Interpretation of the force field parameters is essential in this process. Some properties, including atomic charges, cleavage energy, hydration energy, and facet-specific trends require understanding in the context of chemical knowledge, experimental data, and comparisons to chemically similar compounds across the periodic table to establish validity.[33] Full automation of such cross-checks by genetic algorithms and numerical fitting procedures would compromise the chemical validity and accuracy of the force field. The careful chemical and physical examination of the force field parameters combined with thorough review of the literature on surface chemistry is also quintessential to include features such as pH dependence. This approach even outperforms DFT-level quantum mechanical accuracy in properties such as surface energies and mechanical properties (see Table 6). At the same time, the number of significant digits in the parameter set is small (Tables 1 and 2). Excess digits of parameters, as commonly reported in ab-initio derived force fields, often exaggerate the accuracy and sensitivity of the force field, making it harder for inexperienced users to interpret parameters and try modifications. The philosophy of this work is the exclusion of irrelevant information and focus on a "chemical code" (accurate and minimalist parameters) that help in understanding chemical bonding and nonbonded forces in hydroxyapatite. This force field is addressed towards new users to learn about key aspects of the parameterization, try variations and extensions. One such extension may be, for example, the customization for chemical reactions by transcription of the underlying chemistry in the form of charge redistributions, modified bonding, and the treatment of rate determining steps by added Morse potentials.[34, 35]

**S2.2. Density.** The density was initially computed using a 3D periodic model of a 2×2×2 supercell of HAP.[2] The model was built using multiples of the unit cell based on XRD measurements. Following 50-200 steps of energy minimization, molecular dynamics simulation in the NPT ensemble was carried out for 50 ps for initial relaxation (data discarded), followed by another 50 ps period for recording average cell parameters and thermodynamic properties. The Discover program[36] was employed with a time step of 0.5 fs, Parrinello-Rahman pressure control[37] at 1 atm and temperature control at 298.15 K using velocity scaling with a temperature window of ±10 K. Alternative to velocity scaling, the Andersen thermostat with a collision ratio leading to velocity renormalization every 300+ time steps, or the Berendsen thermostat[38] with a relaxation time of 0.1 ps can be employed. The Parrinello-Rahman barostat allows changes in the shape of the simulation cell, and equilibrium lattice parameters a, b, c, α, β, and γ obtained by NPT simulations were compared to XRD data to test the performance of force field parameters for hydroxyapatite. The accuracy of Ewald summation[39] for Coulomb interactions was set $10^{-4}$ kcal/mol (sufficient for structural analysis) and the cutoff distance for the summation of Lennard-Jones interaction was set to 12 Å. Simulations were also carried out using larger hydroxyapatite supercells up to 10×10×10 nm$^3$ for extended times of 10 ns, showing the same cell parameters within 0.5%. Arithmetic mixing rules were employed for the simulation using the 12-6 Lennard-Jones potential, and Waldman-Hagler combination rules were applied for 9-6 Lennard-Jones potential (PCFF), following standard procedures.

**S2.3. IR and Raman Spectrum.** The superposition of IR and Raman spectra of hydroxyapatite was computed using the equilibrium 3D periodic model of a 2×2×2 supercell after the calculation of the density (section S2.2). The end structure of the equilibrium trajectory was subjected to further 3 ps of molecular dynamics, whereby snapshots were collected every 1 fs. The vibrational frequencies (IR and Raman) were obtained by computing the velocity autocorrelation function (VACF) of all atoms in this 3 ps trajectory, followed by the Fourier transform.

**S2.4. Mechanical Properties.** Bulk moduli $K$, Young's modulus $E$, and the shear modulus $G$ of hydroxyapatite were calculated using MD simulations in the NPT ensemble. Elastic constants were obtained using MD simulations in the NVT ensemble. A 2×2×2 supercell of HAP was relaxed in NPT ensemble at 298.15 K and 1 atm for 1 ns prior to the simulation of mechanical properties to base all subsequent calculations on an equilibrium structure. A supercell with time-averaged, equilibrium cell parameters was employed for follow-on simulations in the NVT ensemble.

In ensuing NPT simulations, individual stresses $\sigma_j$ with $j = 1\ldots 6$ for *xx, yy, zz, xy. xz*, and *yz* components were applied on the material and the strain response $\varepsilon_i$ with $i = 1\ldots 6$ recorded for each stress setting to compute the average moduli Young's moduli $E$ and shear modulus $G$. Thereby, the system was free to change lattice parameters. For the computation of the bulk modulus $K$, the same stress $\sigma$ was applied simultaneously to *xx*, *yy*, and *zz* components. In follow-on NVT simulations, individual strains $\varepsilon_i$ with $i = 1\ldots 6$ were applied onto the supercell and the response stress $\sigma_j$ with $j = 1\ldots 6$ was recorded, respectively, to compute the elastic constants ($C_{ij}$). Thereby, lattice parameters remained fixed in the simulation.

In these computations, the individual tensile (compressive) and shear stresses were varied from 0.1 GPa to 0.9 GPa with an increment of 0.1 GPa. To record the induced strain, the initial structure was relaxed for 50 ps and simulations subsequently continued for 50 ps to compute the average strain in the NPT ensemble. Tensile strain and shear strain were applied from 0.002 to 0.008 with an increment of 0.001. For recording of the stress, the strained structure was relaxed for 50 ps and subsequently simulated for another 50 ps to compute the average stress in the NVT ensemble. The NPT and NVT protocols are otherwise identical as described for the calculation of the density (section S2.2). The Parrinello-Rahman barostat and Berendsen thermostat were applied to control stress and temperature, respectively.

Note that experimental measurements of isotropic moduli were carried out on polycrystalline hydroxyapatite and converted into corresponding isotropic moduli for single crystal hydroxyapatite using the VRH approximation; the difference is typically only ~1% and the error <1% as well.[40]

**S2.5. Cleavage Energy.** Cleavage energies are representative of surface energies for minerals that undergo surface reconstruction upon cleavage. Computed cleavage energies factor in surface reconstruction after cleavage and exclude further modifications such as hydration reactions with water. The computation of cleavage energies was carried out using a two-box method (Figure S1).[31, 32, 41, 42] The two boxes contained a unified and a separated slab of mineral of 6 nm and 3 nm thickness, respectively, and the 3D periodic box contained a slab of vacuum of at least 5 nm above and below the mineral slab in the cleavage direction. Each box had the same lateral dimensions and contained the same total number of atoms in the two-step procedure. The difference in average total energy $\Delta E_{cleav}$ between the separated structure $E_S$ and the unified structure $E_U$ equals the cleavage energy, which was normalized per surface area:

$$\Delta E_{cleav} = \frac{E_S - E_U}{2A} \tag{S3}$$

The surface area equals $2A$ as two surfaces of area $A$ are created upon cleavage. In experiment, cleavage energies $\Delta E_{cleav}$ can be determined by calorimetry.

Alternatively, the free energy of cleavage $\Delta A_{cleav}$ includes additional entropy contributions:

$$\Delta A_{cleav} = \frac{E_S - E_U}{2A} - T\frac{S_S - S_U}{2A} \tag{S4}$$

The entropy contribution $-T(S_S - S_U)/2A$ in equation (S4) is generally small (on the order of -10 mJ/m$^2$) compared to cleavage energy in minerals that ranges from 200 to 1200 mJ/m$^2$ (Figure 5). The small entropy contribution is a result of the restricted mobility of surface molecules and ions resulting from minor oscillations around their lattice positions at the cleaved surface that is only slightly greater in comparison to molecules and ions in the bulk.

The same procedure was also applied to the hydrated and protonated hydroxyapatite surfaces to determine the remaining cleavage energy. These surfaces were first agglomerated and annealed before calculation of the then-remaining cleavage energy upon separation.

Molecular dynamics simulations with the two boxes were conducted in the NVT ensemble at 298.15 K using the Andersen thermostat[43] in Material Studio 4.0 (Figure S1). Ewald summation for Coulomb interactions was carried out with high accuracy of $10^{-6}$ kcal/mol, and the cutoff distance for van der Waals interaction was chosen at 12 Å. The duration of molecular dynamics simulation was 500 ps and the average energies were recorded during the last 100 ps.

**S2.6. Immersion Energy.** Immersion energies were calculated using three simulation boxes, including a solid-water, solid-vacuum, and water-only system (Figure S2). The flexible SPC water model was used for simulations with the 12-6 Lennard-Jones potential (CHARMM-INTERFACE), and the SPC-equivalent water model for simulations with the 9-6 Lennard-Jones potential (PCFF-INTERFACE). The water-only box had similar dimensions in $x$ and $y$ directions as the solid-containing boxes; it contained 1300 water molecules with an approximate size of 30 Å × 30 Å × 40 Å. The surface slab of apatite was the same in box 1 and 2, including the same number and initial positions of atoms, and the number of water molecules was the same in boxes 1 and 3 (Figure S2). Molecular dynamics simulations were carried out in the NPT ensemble for all boxes, whereby simulations of box 2 with vacuum in the z direction were also carried out in the NVT ensemble with identical results using the average dimensions in $x$ and $y$ directions derived from NPT simulation. The simulation time was 1 ns for equilibration and 5 ns for recording average thermodynamic properties using a time step of 1 fs. The summation of pairwise interactions in the Lennard-Jones potential was carried out with a spherical cutoff at 12 Å, and the electrostatic energy was calculated by the Particle-Particle/Particle-Mesh Ewald (PPPM)[44] with high accuracy of $10^{-6}$ kcal/mol. The temperature was kept at 298.15 K and the pressure at 1.0 atm using the Nose-Hoover thermostat and barostat,[45-47] respectively. The program LAMMPS[48] was employed to carry out the simulations. The average energies obtained from the simulations of each box were corrected for the exact temperature of 298.15 K using the heat capacity of the system. This correction was necessary to eliminate errors in the calculation of the immersion energy as a difference in average energies of the three simulation boxes (Figure S2). The resulting numerical uncertainty of the calculated heat of immersion was less than 5%.

**S2.7. Density Profiles in Water.** The density profiles of phosphate ions, calcium ions, and water were computed from the equilibrium trajectories of the apatite-water interfaces for different facets and pH conditions as used in the calculation of immersion energies. Reported densities (Figure 7) rely (1) on the positions of the phosphate atoms to represent the average position of the phosphate, hydrogen phosphate, or dihydrogen phosphate ions, (2) on the positions of the calcium ions, and (3) on the number density of both O and H atoms in water, converted into the corresponding density of water (note that this differs slightly from a density profile based purely on the masses of each atom since hydrogen and oxygen differ in mass). The data are averaged over 500 independent snapshots (~5000 atoms each) over a period of 5 ns. Each of the 36 graphs (Figure 7) relies on over 2 million independent atomic positions and the bin size was chosen as 0.01 Å with 3-point averaging.

**S2.8. Peptide Adsorption.** Simulations were carried out in the NPT ensemble, including the chosen hydroxyapatite surfaces, the peptide S(+)VSVGGK(+)(-)·Cl(-) in zwitterionic state with a chloride ion to compensate the positive charge on the lysine chain, and 1500 water molecules. For each system, five independent start conformations were prepared. In three of the structures, the peptides were initially placed on the surface in vacuum, followed by MD simulation and subsequent addition of water. In two of the start structures, the peptides were initially placed within 0.5 nm distance near the surface after addition of water, followed by minimization to remove atomic close contacts without loss of stereochemistry. A short energy minimization and simulation times of up to 20 ns were then employed for sufficient conformation sampling of the five peptide-containing structures using a time step of 1 fs. The program NAMD was used with pairwise summation of Lennard-Jones interactions with a spherical cutoff at 12 Å, the PME method to compute the electrostatic energy[44] in high accuracy of $10^{-6}$ kcal/mol, the Langevin barostat at 101 kPa and the Langevin thermostat at 298.15 K. All start structures converged towards similar conformations and average energies (+/- 2 kcal/mol in block averages). The final 5 ns of simulation time of the five trajectories were employed to analyze the time-average distance of amino acid residues and to identify functional groups in close contact with the surface (Figure 8). The brief results here illustrate the feasibility of simulations of apatite-biomolecular interfaces; complete details will be reported separately. Equivalent simulation protocols were previously described to study binding of peptides to surfaces of silica and metal nanoparticles.[30, 49]

## Supporting References